\documentclass[oneside,twocolumn]{article}

\usepackage{aas}
\usepackage{graphicx}
\usepackage{natbib}
\usepackage{authblk}
\usepackage{flafter}
\usepackage[version=3]{mhchem} 
\usepackage{siunitx} 
\usepackage{mathabx} 
\usepackage{amsmath}
\usepackage[table]{xcolor}
\usepackage[colorlinks,bookmarks=false,linkcolor=blue,urlcolor=red,citecolor=blue]{hyperref}
\usepackage{geometry}
\usepackage{times}

\DeclareSIUnit\year{yr}
\DeclareSIUnit\parsec{pc}
\DeclareSIUnit{\wtpercent}{wt\%}

\newcommand{\vplanet}{\texttt{\footnotesize{VPLanet}}}
\newcommand{\atmesc}{\texttt{\footnotesize{AtmEsc}}}

\newcommand{\eqtide}{\texttt{\footnotesize{EqTide}}}

\newcommand{\radheat}{\texttt{\footnotesize{RadHeat}}}

\newcommand{\stellar}{\texttt{\footnotesize{STELLAR}}}

\newcommand{\magmoc}{\texttt{\footnotesize{MagmOc}}}

\newcommand{\petit}{\textit{petitCODE}}
\usepackage{txfonts}
\usepackage[shortcuts]{extdash}
\newcommand{\eg}{e.g. }

\title{Magma ocean evolution of the TRAPPIST\=/1 planets}

\author{Patrick Barth$^{1,2,3}$, Ludmila Carone$^{3}$, Rory Barnes$^{4,5}$, Lena Noack$^{6}$, Paul Molli\`{e}re$^{3}$, and Thomas Henning$^{3}$}

\date{}

\begin{document} 
\twocolumn[
\begin{@twocolumnfalse}
\maketitle
$^1$Centre for Exoplanet Science, University of St Andrews, North Haugh, St Andrews, KY169SS, UK \\
$^2$SUPA, School of Physics \& Astronomy, University of St Andrews, North Haugh, St Andrews, KY169SS, UK \\
$^3$Max Planck Institute for Astronomy, K\"onigstuhl 17, 69117 Heidelberg, Germany \\
$^4$Astronomy Department, University of Washington, Box 3515580, Seattle, WA 98195, USA \\
$^5$NASA Virtual Planetary Laboratory Lead Team, USA \\
$^6$Freie Universit\"at Berlin, Institute of Geological Sciences, Malteserstr. 74-100, 12249 Berlin, Germany \\

\textbf{Keywords:} Exoplanets, Terrestrial Planets, Planetary Atmospheres, Magma oceans

\textbf{Corresponding Author:} Patrick Barth, School of Physics and Astronomy, University of St Andrews, KY169SS, UK, \\
pb94@st-andrews.ac.uk

\date{Accepted for publication in Astrobiology on May 10, 2021}

\begin{abstract}

Recent observations of the potentially habitable planets TRAPPIST\=/1~e, f, and g suggest that they possess large water mass fractions of possibly several tens of wt\% of water, even though the host star's activity should drive rapid atmospheric escape. These processes can photolyze water, generating free oxygen and possibly desiccating the planet. After the planets formed, their mantles were likely completely molten with volatiles dissolving and exsolving from the melt.
In order to understand these planets and prepare for future observations, the magma ocean phase of these worlds must be understood. To simulate these planets, we have combined existing models of stellar evolution, atmospheric escape, tidal heating, radiogenic heating, magma ocean cooling, planetary radiation, and water-oxygen-iron geochemistry.
We present \magmoc{}, a versatile magma ocean evolution model, validated against the rocky Super-Earth GJ~1132b and early Earth. We simulate the coupled magma ocean\=/atmospheric evolution of TRAPPIST\=/1~e, f, and g for a range of tidal and radiogenic heating rates, as well as initial water contents between 1 and 100 Earth oceans. 
We also reanalyze the structures of these planets and find they have water mass fractions of 0--0.23, 0.01--0.21, and 0.11--0.24 for planets e, f, and g, respectively.
Our model does not make a strong prediction about the water and oxygen content of the atmosphere of TRAPPIST\=/1~e at the time of mantle solidification. 
In contrast, the model predicts that TRAPPIST\=/1~f and g would have a thick steam atmosphere with a small amount of oxygen at that stage.
For all planets that we investigated, we find that only $3-5\%$ of the initial water will be locked in the mantle after the magma ocean solidified.
   
\end{abstract}
\end{@twocolumnfalse}]

\section{Introduction}
As of March 2021, there are 4692 confirmed exoplanets\footnote{\url{http://exoplanet.eu/catalog/}}.
Of those exoplanets, 60 are optimistically classified as potentially habitable\footnote{\url{http://phl.upr.edu/projects/habitable-exoplanets-catalog}}, meaning they orbit in the star's habitable zone and are rocky. The habitable zone is defined as the region around a star where liquid water can be present on the surface of a rocky planet, assuming a greenhouse atmosphere \citep{Huang1959, Kasting1993, Kopparapu2013}. According to population statistics from the Kepler exoplanetary survey, a planet should likely be less massive than $5 \, M_\Earth$ \citep{Otegi2020} and smaller than $1.7 R_{\Earth}$ \citep[\eg][]{Ginzburg2018} to be rocky.
Only 14 of these 60 planets are in the conservative habitable zone and have a mass smaller than $5 \, M_\Earth$ and are, therefore, currently the most likely candidates for being habitable rocky worlds.
Here, we use the definition of the conservative habitable zone by \citet{Kopparapu2014} with the inner edge being the mass-dependent runaway greenhouse limit and the outer edge the maximum greenhouse limit.
However, it is important to note that a planet's position in the habitable zone is only one of many factors that determine the habitability of said planet.

Most of these potentially habitable planets are orbiting M dwarfs, including the planets closest to Earth, \eg Proxima Centauri b \citep{Anglada-Escude2016}, TRAPPIST-1 e, f, \& g \citep{Gillon2016}, and Teegarden's Star c \citep{Zechmeister2019}, which are prime candidates for observations with upcoming space- and ground-based instruments like the \textit{James Webb Space Telescope (JWST)} \citep{Barstow2016, Snellen2017, Lincowski2018} or the \textit{Extremely Large Telescope (ELT)} \citep{Snellen2015, Meadows2017}.
To define the goals for these measurements and identify detectable tracers for life, a firm understanding of the evolution of terrestrial planets around M dwarfs is required \citep{Meadows2017b, Catling2018}.

Planets orbiting M dwarfs are in a very different environment than Earth.
These stars are active and a large number of flares reach these rocky planets, which orbit their stars on a much tighter orbit than Earth orbits the Sun \citep{Vida2017a, Gunther2020}. The intense X-ray and UV irradiation (XUV) emitted by active M dwarfs leads to enhanced atmospheric erosion \citep{Watson1981, Lammer2003, Lammer2009, Erkaev2007, Owen2012} that could harm organisms living on the surface of the planet \citep{Lammer2007}.
During the pre-main sequence phase, M dwarfs are much more luminous, with their potentially habitable planets receiving XUV fluxes that are up to 100 times more powerful than on the modern Earth \citep{Lammer2007}.

Studies on the composition of the TRAPPIST\=/1 planets by \citet{Dorn2018} and \citet{Unterborn2018} suggest that some of these planets today have water mass fractions of possibly several tens of wt\%, but with large uncertainties. These abundances are orders of magnitude higher than the estimated current water mass fraction of Earth (0.07 - 0.23 wt\% water or three to ten ``terrestrial oceans\footnote{One terrestrial ocean (TO) is the equivalent of all the water in Earth's oceans ($1.39 \times 10^{21} \si{\kilogram}$)},'' \eg \citet{Elkins-Tanton2008,Pearson2014,Schmandt1265}). This discrepancy suggests a very different formation and evolution history for the TRAPPIST\=/1 planets compared to the rocky Solar System planets, which is still the subject of ongoing research.
Therefore, we need new models to simulate the evolution of terrestrial planets around M dwarfs to be able to explain how much water the TRAPPIST\=/1 planets could have held on to during their evolution and how much water they must have initially received during their formation in the first place.

Due to the heat produced by accretion, core formation, as well as radiogenic and tidal heating, most planetary-sized bodies in the Solar System appear to have been in a fully molten state right after formation, during which the mantle vigorously convected while gases passed between the atmosphere and liquid rock with relative ease \citep{Wood1970, Solomon1979, Wetherill1990, Lammer2018A}. 
This planetary state, called a magma ocean, drives rapid geochemical evolution that dramatically alters the redox state of the mantle and atmosphere. 
A thick greenhouse atmosphere heated by the surface and stellar irradiation moderates the cooling of the magma ocean. 
In many cases, an active nearby star releases high energy photons that dissociate water and remove the atmosphere. Hydrogen escapes more easily than oxygen, so the latter will either enter the melt and bond with iron, or accumulate in the atmosphere \citep[\eg][]{Luger2015a}.
This conceptual model of a magma ocean, which is greatly simplified, is shown schematically in Fig.~\ref{struc_magma}. Note that this is assuming an oxidized magma ocean which is outgassing \ce{H2O} rather than \ce{H2}.

Such a scenario was tackled with simulations for rocky Solar System planets initially for an early atmosphere dominated by water vapour by \citet{Matsui1986,Zahnle1988,Abe1993,Abe1997,Solomatov2000,Zahnle2015} and by \citet{Elkins-Tanton2008} for an early $\mathrm{H_2O}$ and $\mathrm{CO_2}$ atmosphere.
\citet{Schaefer2016} were the first to investigate magma ocean outgassing and atmosphere erosion for a rocky exoplanet around an M dwarf star, the hot, un-inhabitable Super-Earth GJ~1132b.
They found that around a pre-main sequence M dwarf, atmosphere erosion can be so effective that abiotic free $\mathrm{O_2}$ builds up in the early atmosphere, as much as several thousands of bars. 

Abiotic oxygen build-up on rocky planets around M dwarfs was already proposed by \citet{Luger2015a}, but without taking into account oxygen solubility back into the magma ocean. \Citet{Schaefer2016} considered a pure steam atmosphere and did not take into account the contribution of $\mathrm{CO_2}$. They also did not take into account that additional heating sources like tidal heating \citep[\eg][]{Driscoll2015} and magnetic induction \citep{Kislyakova2017} could potentially delay solidification of the magma ocean and thus reduce abiotic oxygen build-up.

In this study, we consider magma ocean volatile outgassing and atmosphere erosion for the potentially habitable \mbox{TRAPPIST\=/1} planets e, f, g considering an early \ce{H2O} dominated atmosphere. We consider initial water contents from 1 - 100 terrestrial oceans and compare the final water content to the results obtained by different interior structure models \citep{Noack2016,barr2018interior,Dorn2018,Unterborn2018b}. We also investigate how different levels of interior heating, from radiogenic and tidal sources, affect the final atmospheric composition.

To perform our simulations, we have augmented the open source software package \vplanet{} \citep{Barnes2020} to include the physics and chemistry necessary to model the TRAPPIST\=/1 planets during the magma ocean phase. \vplanet{} already includes ``modules'' for radiogenic heating (\radheat{}), tidal heating (\eqtide{}), stellar evolution (\stellar{}) and water photolysis and hydrogen escape (\atmesc{}). Here we introduce a new module \magmoc{} that calculates the thermal evolution of a magma ocean, volatile fluxes through the surface, and radiative cooling through an \ce{H2O} atmosphere.

In Section \ref{Sec_Model}, we present the magma ocean model, which builds from \citet{Schaefer2016} and \citet{Elkins-Tanton2008}. We further describe different calculations of the atmospheric net flux that leads to the cooling of the planet. 
We discuss our model in comparison to previous work by \citet{Schaefer2016} for GJ~1132b and \citet{Elkins-Tanton2008} and \citet{Hamano2013} for Earth in Section \ref{chap_validation}.
In Section \ref{chap_results}, we present the results obtained for TRAPPIST\=/1~e, f, and g. We discuss the ramification of our results for the TRAPPIST\=/1 planets in Section~\ref{sec_discuss}.
We summarize our results in Section~\ref{sec: conclusion} and conclude with an outlook, where we offer our magma ocean model to the community as the \magmoc{} module of \vplanet{}  (Section~\ref{sec: outlook}).

\section{The Model}
\label{Sec_Model}

In this section, we describe the set-up of the \magmoc{} module. We stress that we describe here a relatively simple initial set-up to establish an open-source magma ocean model that is useful for application to extrasolar and Solar System rocky planets and for a broad community of researchers. In the near future, this model will be expanded to include more complex descriptions of the magma ocean.

\subsection{Thermal Model}
\label{sec_therm_model}
\begin{figure}
    \centering
    \includegraphics[width=\columnwidth]{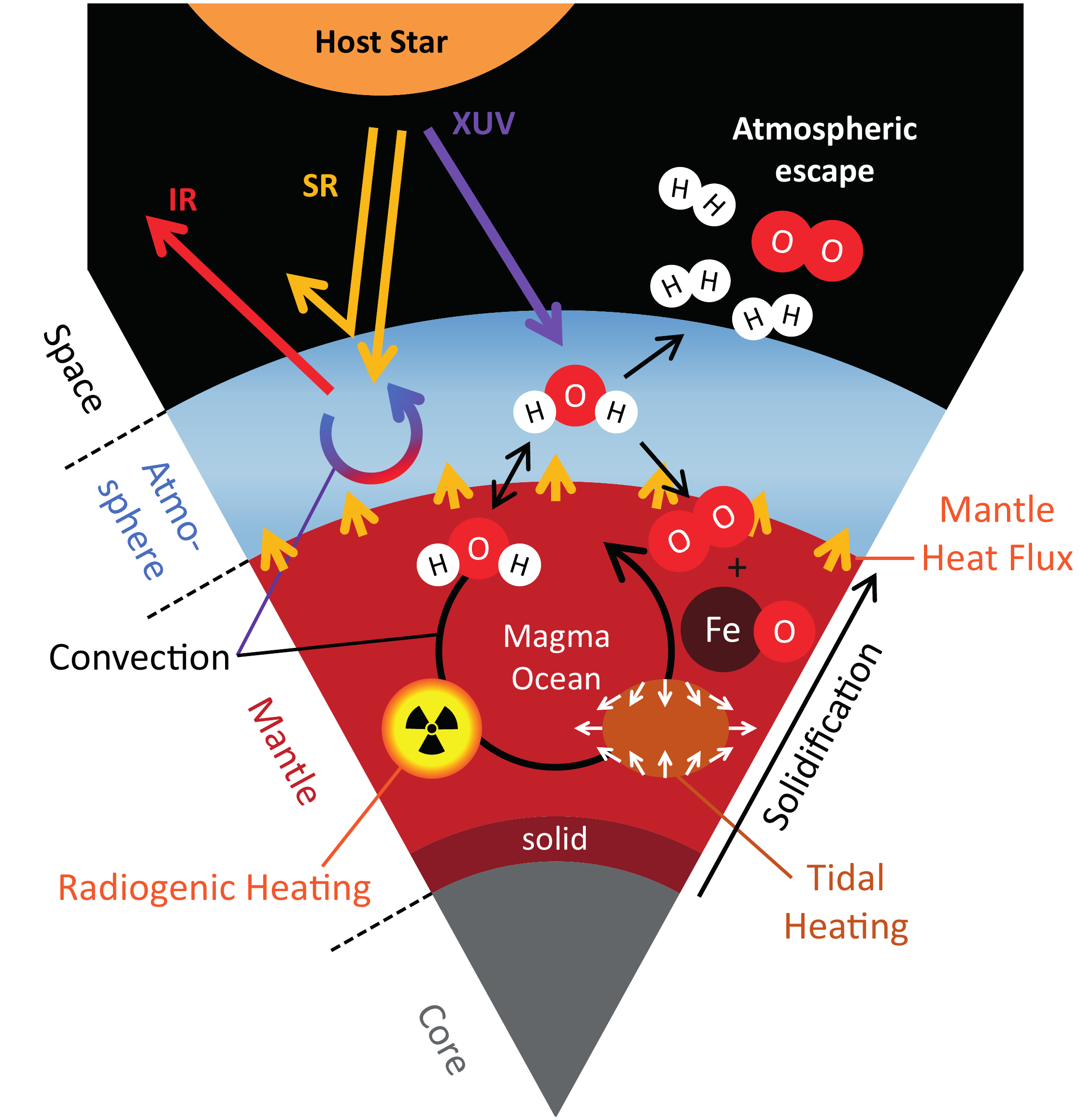}
    \caption{Structure of the magma ocean and atmosphere. Physical and chemical processes between the different planetary layers and the star are shown.}
    \label{struc_magma}
\end{figure}

\begin{table}[ht]
    \caption[Parameters thermal model]{Parameters for the thermal model}
    \begin{tabular}{cl}
    	\noalign{\smallskip}
    	\hline
    	\noalign{\smallskip}
    	Symbol & Parameter \\ 
    	\noalign{\smallskip}
    	\hline \hline
    	\noalign{\smallskip}
    	$T_\mathrm{p}$ & Potential temperature of the mantle \\
    	$T_\mathrm{surf}$ & Surface temperature \\
    	$T_\mathrm{solidus}$ & Solidus temperature \\
    	$T_\mathrm{liquidus}$ & Liquidus temperature \\
    	$r_\mathrm{s}$ & Solidification radius \\
    	$\rho_\mathrm{m}$ & Mantle bulk density \\ 
    	$c_p $ & Silicate heat capacity \\
    	$\Delta H_\mathrm{f}$ & Heat of silicate fusion \\
    	$Q_\mathrm{r}$ & Heat by radioactive decay $[\si{\watt\per\kilogram}]$\\
    	$Q_\mathrm{t}$ & Heat by tidal interactions${}^{a}$ $[\si{\watt\per\kilogram}]$\\
    	$F$ & Heat flux leaving atmosphere (OLR - ASR ${}^{b}$)\\
    	$\alpha$ & Thermal expansion coefficient \\
    	$\eta$ & Dynamic viscosity $[\si{\pascal\second}]$\\
    	$\psi$ & magma ocean averaged melt fraction \\
    	$\psi_\mathrm{surf}$ & surface melt fraction \\
    	\noalign{\smallskip}
    	\hline
    \end{tabular}
    \\
    ${}^{a}$ not included in \citet{Schaefer2016}\\
    ${}^{b}$ OLR = Outgoing Long-wave Radiation, ASR = Absorbed Stellar Radiation
    \label{Tab_Therm_Model}
\end{table}

Our thermal model is based in large parts on the coupled stellar-atmosphere-interior model for terrestrial exoplanets by \citet{Schaefer2016} and for rocky Solar System planets by \citet{Elkins-Tanton2008}. If not otherwise noted, the equations in this section are taken from \citet{Schaefer2016}. A concise list of the most important parameters in the thermal model is provided in Table~\ref{Tab_Therm_Model}. Their values, if appropriate, used for our simulations can be found in the appendix in Table~\ref{Tab_Therm_Model_Value}. If not otherwise noted, these values are adapted from \citet{Schaefer2016}.

We start our simulations always with a completely molten mantle, which solidifies from the base to the top. In addition to the primordial heat from accretion, the magma ocean is heated by the decay of radioactive isotopes and tidal interactions. Tidal heating was not considered by \citet{Schaefer2016} even though it can be a very large source of energy \citep{Barnes2010,Barnes2013,Driscoll2015}. The magma ocean cools by turbulently convecting internal energy to the surface, where it is ultimately radiated into space. The atmosphere, in our case a thick atmosphere composed of water, is convective in itself. It cools by emitting infrared radiation (IR) into space, while it is heated by stellar radiation (SR). Part of SR is reflected back to space due to the albedo of a water vapour atmosphere. A schematic of such a planet and the processes in our model is shown in Fig.~\ref{struc_magma}\footnote{There appears to be some inconsistency in the use of the term ``magma ocean'' in the literature. Here, we include the part of the mantle above the radius of solidification $r_\mathrm{s}$ where the melt fraction reaches $\psi = 0$.}.

For the age of the star at the beginning of our simulations, we use the estimated lifetime of the protoplanetary disk, since we are not taking into account further impacts during the magma ocean evolution.
\citet{Ribas2014} give an estimate for the lifetime of $4.2 - \SI{5.8}{\mega\year}$ for disks of low-mass stars, based on statistics of the solar neighborhood.
Therefore, we use an age of the star of $\SI{5}{\mega\year}$ at the beginning of our simulations.

The heating and cooling processes of the planet are controlled by the thermal model which tracks the potential temperature $T_\mathrm{p}$ of the mantle and the solidification radius $r_\mathrm{s}$ for every time step.
$T_\mathrm{p}$ is the temperature of the uppermost layer of the magma ocean and defines the temperature profile of the mantle (see Eq.~\ref{adiabatic}). 
As convection in the molten part of the mantle is very efficient, the magma ocean and surface are in thermal equilibrium (see Sec.~\ref{subsec_convec}).
We therefore assume the surface temperature to be $T_\mathrm{surf} = T_\mathrm{p}$.
In addition, the thermal model keeps track of the melt fraction $\psi$ and the time derivative of $r_\mathrm{s}$ which are needed for the volatile model. 
The initial parameters are $T_\mathrm{p} = \SI{4000}{\kelvin}$ and $r_\mathrm{s} = r_\mathrm{c}$, where $r_\mathrm{c}$ is the core radius.
This temperature was chosen to ensure that the mantle is completely molten at the start of the simulation.
(See Section~\ref{SubSubSec_Solid} for more information on the treatment of the solidification radius.)

More precisely, the magma ocean temperature is described by the following equation \citep[adapted from][]{Schaefer2016}:
\begin{multline}
	\label{Deriv_Tpot}
	\frac{4}{3} \pi \rho_\mathrm{m} c_p (r^3_\mathrm{p} - r^3_\mathrm{c}) \epsilon_\mathrm{m} \frac{dT_\mathrm{p}}{dt} = 4 \pi r^2_\mathrm{s} \Delta H_\mathrm{f} \rho_\mathrm{m} \frac{dr_\mathrm{s}}{dt} \\ 
	+ \frac{4}{3} \pi \rho_\mathrm{m} (Q_\mathrm{r} + Q_\mathrm{t}) (r^3_\mathrm{p} - r^3_\mathrm{c}) - 4 \pi r^2_\mathrm{p} F
\end{multline}
with the first term on the right hand side describing the heat produced by the fusion of silicates (latent heat), the second term the radiogenic and tidal heating and the third term is the atmospheric net flux cooling the planet off to space.
This equation is based on the model by \citet[Eq.~2]{Hamano2013} but includes additional heat sources on the right hand side that are also partly included in \citet{Schaefer2016}.
For simplification of the model we assume the additional heat by radiogenic and tidal heating is distributed over the whole mantle and not just the magma ocean.
Therefore, we apply this equation on the whole mantle as well, i.e. the average mantle temperature $T_\mathrm{m}$.
This temperature is related to the potential temperature by $T_\mathrm{m} = \epsilon_\mathrm{m} T_\mathrm{p}$ with $\epsilon_\mathrm{m} = 1.19$ for planets of approximately one Earth mass \citep{Schaefer2015}.
Following previous models \citep{Elkins-Tanton2008,Hamano2013,Lebrun2013,Schaefer2016,Nikolaou2019} we do not include the heat flux from the core into the mantle.
In the following, we describe the terms on the right hand side of Eq.~(\ref{Deriv_Tpot}) that describes the thermal evolution of the planet.

\subsubsection{Parameterization of convection}
\label{subsec_convec}

The cooling and solidification of the mantle is driven by very strong convection due to the low viscosity of the liquid melt.
Only when the surface melt fraction drops below a critical value $\psi_\mathrm{c}$ does the viscosity increase.
The viscosity thus depends on the melt fraction $\psi$. 
The melt fraction at the surface is expressed for $T_\mathrm{solidus} < T_\mathrm{p} < T_\mathrm{liquidus}$ by
\begin{equation}
    \psi_\mathrm{surf} = \frac{(T_\mathrm{p} - T_\mathrm{solidus,surf}) }{ ( T_\mathrm{liquidus,surf} -  T_\mathrm{solidus,surf}) },
\end{equation}
with $T_\mathrm{solidus,surf} = \SI{1420}{\kelvin}$ and $T_\mathrm{liquidus,surf} = \SI{2020}{\kelvin}$.

\citet{Lebrun2013} showed that $\psi = 0.4$ marks the phase transition towards a thicker, viscous thermal boundary layer and the effective end of the magma ocean stage\footnote{\citet{Lebrun2013} define the location in the mantle where $\psi = 0.4$ as the rheology front $r_\mathrm{f}$}.
For the viscosity profile, therefore, two different regimes of the surface melt fraction are considered:
Above the critical melt fraction $\psi_\mathrm{c} = 0.4$, the viscosity parameterization from \citet{Lebrun2013} is used, while for $\psi_\mathrm{surf} < 0.4$, the solid-like viscosity needs to be applied and we follow \citet{Lebrun2013} and \citet{Schaefer2016}.
The dynamic viscosity $\eta$ is thus defined by:
\begin{equation}
\label{eta_liq}
    \eta = 
        \begin{cases}
        \eta_\mathrm{l} \left( 1 - \frac{1-\psi}{1-\psi_\mathrm{c}}\right)^{-2.5} 
        \exp \left( \frac{\SI{4600}{\kelvin}}{T_\mathrm{p} - \SI{1000}{\kelvin}} \right)
        & \text{if $\psi_\mathrm{surf} \ge \psi_\mathrm{c}$} \\
        \eta_\mathrm{s} \exp \left( \frac{E_a}{R T_\mathrm{p}} \right)  
        & \text{if $\psi_\mathrm{surf} < \psi_\mathrm{c}$}
        \end{cases},
\end{equation}
with $\eta_\mathrm{l} = \SI{0.00024}{\pascal\second}$, $\eta_\mathrm{s} = 3.8 \times 10^9 \si{\pascal\second}$, the activation energy $E_a = \SI{350}{\kilo\joule\per\mole}$, and the ideal gas constant $R$.

This returns a viscosity of $\eta < \SI{0.1}{\pascal\second}$ for a fully molten surface and only for a surface melt fraction of just above 0.4 the viscosity increases to $\eta > 10^3 \si{\pascal\second}$.
We therefore conclude that our approximation of efficient convection is valid for most of the solidification process\footnote{See Sec.~\ref{sec_discuss} for a discussion of the final 2\% of mantle solidification.}.
Please note that we did not include the effect of water in the melt on the viscosity of the magma ocean.
Studies by \citet{Nikolaou2019} show that the effect of water dependent changes of the viscosity on the solidification time of the mantle is negligible.

\subsubsection{Mantle solidification}
\label{SubSubSec_Solid}

To calculate changes in the solidification radius $r_\mathrm{s}$, we define the adiabatic temperature structure $T(r)$ of the convective mantle with respect to the potential temperature $T_\mathrm{p}$:
\begin{equation}
\label{adiabatic}
    T(r) = T_\mathrm{p} \lbrack 1 + \frac{\alpha g}{c_p} (r_\mathrm{p} - r) \rbrack.
\end{equation}
Here, $g$ is the planetary surface gravity, $c_p$ is the specific heat capacity of the magma ocean and $\alpha$ is the thermal expansion coefficient.

To parameterize the solidification, we use the solidus and liquidus profiles from \citet{Hirschmann2000}:
\begin{equation}
\label{eq_solidus}
    T_\mathrm{solidus}(p) = a p + b
\end{equation}
where $a = \SI{104.42}{\kelvin\per\giga\pascal}$ and $b = \SI{1420}{\kelvin}$ for ${p < \SI{5.2}{\giga\pascal}}$, and $a = \SI{26.53}{\kelvin\per\giga\pascal}$ and $b = \SI{1825}{\kelvin}$ for ${p > \SI{5.2}{\giga\pascal}}$.
The depth $z$ dependent pressure is given by $p(z) = \rho_\mathrm{m} g (z - z^2 / 2 r_\mathrm{p})$.
The liquidus temperature is assumed to be larger than the solidus by $\SI{600}{\kelvin}$.
The solidification radius $r_\mathrm{s}$ is defined as the radius where the mantle temperature $T$ equals $T_\mathrm{solidus}$ where the melt fraction is $\psi = 0$.

Equations~(\ref{adiabatic}) and (\ref{eq_solidus}) are used to calculate the change of the solidification radius $r_\mathrm{s}$:
\begin{equation}
	\label{Deriv_rsol}
	\frac{dr_\mathrm{s}}{dt} = \frac{c_p (b \alpha - a \rho_\mathrm{m} c_p)}{g (a \rho_\mathrm{m} c_p - \alpha T_\mathrm{p})^2} \frac{dT_\mathrm{p}}{dt}.
\end{equation}

We caution the reader that $r_\mathrm{s}$ only changes when the temperature at the core-mantle boundary $r_\mathrm{c}$ drops below $T_\mathrm{solidus}$.
When choosing initial conditions such that the whole mantle is molten, i.e. $\psi(r_\mathrm{c})=1$, we set $r_\mathrm{s} = r_\mathrm{c}$ until the base of the mantle starts to solidify.

\subsubsection{Radiogenic and tidal heating}

Radiogenic heat in the Earth is generated by the decay of \ce{^{238}U}, \ce{^{235}U}, \ce{^{232}Th}, and \ce{^{40}K} in the core, mantle, and crust.
The radiogenic power produced by species $i$ in reservoir $j$ is
\begin{equation}
	Q_{i,j}=Q_{i,j}(0)\exp(-\lambda_{i,1/2}t)
	\label{radiogenicheat}
\end{equation}
where $\lambda_{i,1/2}=\ln 2/\tau_{i,1/2}$, $\tau_{i,1/2}$ is the halflife, t is time, and $Q_{i,j}(0)$ is the initial heat production at $t=0$.
This heat source is included in our model by connecting it to the \radheat{} module of \vplanet{} \citep{Barnes2020}. Although the planets may form quickly enough for $^{26}$Al heating to be important \citep{Barnes2016,Lichtenberg2019}, we ignore this effect here\footnote{For the purpose of this study we assume that the start point of our simulation is late enough for $^{26}$Al heating to be negligible}.

Similarly, the tidal heating rate $Q_\mathrm{t}$ is calculated by the module \eqtide{} of \vplanet{}.
It is based on the equilibrium tide theory \citep{Darwin1880,Ferraz-Mello2008,Leconte2010}, assuming the gravitational potential of the tide raiser on an unperturbed spherical surface can be expressed as the sum of surface waves and that the elongated equilibrium shape of the perturbed body is slightly misaligned with respect to the line that connects the two centers of mass \citep{Barnes2020}.
The advantage of this model is that it is semi-analytic and reduces the tidal effects to a single parameter (the tidal quaility factor $Q$), which is especially valuable for exoplanets where little is known about their interior composition.
Self-consistent models, however, would need three-dimensional models including the rheology of the interior.
Ocean worlds would even need three-dimensional models of the ocean currents \citep{Carone2012}.
In our simulations we use the ``constant-phase-lag'' (CPL) model in \eqtide{} where, regardless of orbital and rotational frequencies, the angle between the perturber and the tidal bulge remains constant.
When the body is assumed to behave like a harmonic oscillator, the damping must be independent of the frequency for the tidal waves to be linearly summed.
The lag between the line connecting the two centers of mass and the direction of the tidal bulge is inversely proportional to the ``tidal quality factor'' $Q$.

In this study, we only include the tidal effect on the planet by the star. 
However, in multi-planet systems tidal interactions between individual planets might be important \citep{Wright2018,Hay2019}.
Further, we did not include tidal effects raised by a moon because close in planets are unlikely to have moons due to tidal effects between the star, planet, and moon \citep{BarnesOBrien02,SasakiBarnes14}.

\subsubsection{Atmospheric Flux}
\label{Sec_Flux}

The simplest way to calculate the outgoing long-wave radiation is by implementing a grey atmosphere model. We base this part of our model on the the grey atmosphere model of \citet{Elkins-Tanton2008}, which was applied to Earth, and on the model of \citet{Carone2014}, developed for different atmosphere compositions on rocky exoplanets. In the grey radiative transfer approach, the opacity is not calculated for every wavelength bin. Instead, the average infrared opacities of the individual species are combined, weighted by the partial pressure of the specific gas.

The atmospheric net flux is then calculated by multiplying the difference between the Stefan-Boltzmann emission from the surface ($\sigma T_\mathrm{surf}^4$) and the top of the atmosphere ($\sigma T_\mathrm{eq}^4$) with the emissivity $\epsilon$:
\begin{equation}
\label{Eq_fluxgrey}
F = \epsilon \sigma ( T_\mathrm{surf}^4 - T_\mathrm{eq}^4 ).
\end{equation}
$T_\mathrm{eq}$ is the equilibrium temperature of the atmosphere due to the stellar irradiation.
The emissivity is calculated with the optical depth $\tau^*$ which is the sum of the optical depths of all atmospheric species.
The emissivity $\epsilon$ is given by
\begin{equation}
\epsilon = \frac{2}{\tau^* + 2}
\end{equation}
with the optical depth of the atmospheric species $i$
\begin{equation}
\tau^*_i = \left( \frac{3 M_{i}^\mathrm{atm}}{8 \pi R^2} \right) \left( \frac{k_0 g}{3 p_0} \right)^{1/2}.
\end{equation}
$M_{i}^\mathrm{atm}$ is the atmospheric mass of species $i$.
To calculate the optical depth, we use the atmospheric absorption coefficient $k_{0,\ce{H2O}} = \SI{0.01}{\metre\squared\per\kilogram}$ for water \citep{Elkins-Tanton2008} at the reference pressure $p_0 = \SI{101325}{\pascal}$ \citep{Yamamoto1952}.

One aspect of atmospheric behavior that is not included in the grey atmosphere model is the runaway greenhouse or blanketing effect. This effect prevents the planet from emitting more than $\SI{280}{\watt\per\square\metre}$ when greenhouse gasses in the atmosphere are present, unless the surface temperature is larger than $\SI{1800}{\kelvin}$ \citep{Goldblatt2013, Kopparapu2013}.
Once the surface temperature of a rocky planet with a greenhouse atmosphere enters the runaway greenhouse range ($600-\SI{1800}{\kelvin}$), the planet will ultimately heat up because it cannot emit enough radiation to cool.
For surface temperatures of $T_\mathrm{surf} \lesssim \SI{1600}{\kelvin}$ the temperature of the atmospheric layer at which the optical depth is unity ($\tau \sim 1$) is between $250$ and $\SI{300}{\kelvin}$, which limits the emitted radiation to the Planck functions of those temperatures.
This limit of $\sim \SI{280}{\watt\per\square\metre}$ is called the Simpson-Nakajima radiation limit \citep{Simpson1928, Nakajima1992}.
When the surface temperature exceeds $1600 - \SI{1800}{\kelvin}$, the temperature of the emitting layer ($\tau \sim 1$) reaches $\SI{400}{\kelvin}$ in the $\SI{4}{\micro\metre}$ water vapor window.
The new emerging emission peak allows the planet to emit more than $\SI{280}{\watt\per\square\metre}$ and the planet cools efficiently again.

We treat this effect by adapting the grey atmosphere model to set the outgoing flux to $\SI{280}{\watt\per\square\metre}$ if the surface temperature is in the range of $600-\SI{1800}{\kelvin}$ and the water pressure is greater than $\SI{10}{\milli\bar}$.
If the absorbed stellar radiation is larger than this flux, the surface temperature will stay constant at $\sim \SI{1800}{\kelvin}$ until the atmosphere is desiccated due to the atmospheric escape of hydrogen or the absorbed stellar radiation drops below the runaway greenhouse limit.

To benchmark our adapted grey atmosphere model, we used the \petit{} \citep{Molliere2015,Molliere2017} to calculate the net outgoing flux from a pure steam atmosphere.
The \petit{} is a one-dimensional planetary atmosphere code that calculates the vertical structure of an optically thick (at the bottom) planetary atmosphere using the correlated-k method.
It includes condensation and cloud formation as well as an equilibrium chemistry model for a wide range of atmospheric compositions.

We ran simulations for a range of surface temperatures and atmosphere pressures and calculated the Outgoing Longwave Radiation (OLR) for a 100\% water vapor atmosphere on the extrasolar Super-Earth GJ~1132b \citep{Berta-Thompson2015}. Based on this, we created an interpolated look-up table for a given surface temperature and atmospheric pressure to calculate the OLR during runtime and implemented it into our model.

In Section \ref{sec_res_GJ}, we show the magma ocean evolution on GJ~1132b for both radiative transfer models (grey \& \petit{}) and compare the results to those from \citet{Schaefer2016} who used a line-by-line radiative transfer model.

\subsection{Volatile Model}
\label{sec_vol_model}

\begin{table}[ht]
	\caption[Parameters volatile model]{Parameters for the volatile model}
	\begin{tabular}{cl}
		\noalign{\smallskip}
		\hline
		\noalign{\smallskip}
		Symbol & Parameter \\ 
		\noalign{\smallskip}
		\hline \hline
		\noalign{\smallskip}
		$F_{\ce{H2O}}$ & Water mass fraction in liquid melt \\ 
		$M_{\ce{H2O}}^{\mathrm{mo}} $ & Water mass magma ocean + atmosphere \\
		$M_{\ce{H2O}}^{\mathrm{cyrstal}} $ & Water mass crystals in magma ocean \\
		$M_{\ce{H2O}}^{\mathrm{liq}} $ & Water mass in the liquid melt \\
		$M_{\ce{H2O}}^{\mathrm{atm}} $ & Water mass in the atmosphere \\
		$M_{\ce{H2O}}^{\mathrm{sol}} $ & Water mass in the solidified mantle \\
		$k_{\ce{H2O}}$ & Water part. coeff. melt - solid \\
		$\phi_1$ & XUV-driven atm. mass-loss rate of H $[\si{kg\per\metre\squared\per\second}]$\\
		$\phi_2$ & XUV-driven atm. mass-loss rate of O $[\si{kg\per\metre\squared\per\second}]$\\
		\noalign{\smallskip}
		\hline
	\end{tabular}
	\label{Tab_Volat_Model}
\end{table}

The volatile model controls the exchange of water and oxygen between the different layers of the system.
Unlike for a cool planet with a solid surface, the partitioning of volatiles between the interior and atmosphere are in equilibrium during a magma ocean phase. Most importantly, the mass fraction of water in the melt is set by the partial pressure of water in the atmosphere. Therefore, the system is characterized by, and very sensitive to, sources and sinks.

At the base of the magma ocean, melt solidifies due to the cooling of the planet, locking water in the solid rock. However, the solid rock can only keep a small part of the water that was dissolved in the melt. The majority of the water stays in the melt increasing the mass fraction of water and, therefore, water is outgassed into the atmosphere to increase the pressure and restore the equilibrium.

At the top of the atmosphere water is lost into space because the star emits XUV radiation which dissociates water in the atmosphere into hydrogen and oxygen.
While the hydrogen and part of the oxygen escape to space, the rest of the oxygen stays in the atmosphere and dissolves into the magma ocean where it oxidizes \ce{FeO} to \ce{Fe2O3}.
When all of the \ce{FeO} in the melt is oxidized or the mantle has solidified completely, oxygen starts to build up in the atmosphere.

The volatile model uses the thermal model to calculate the amount of water and oxygen that is stored in the atmosphere + magma ocean system (MOA) and the solid mantle (SOL). We assume that in the beginning the MOA contains all the water and no free oxygen is present.
In equilibrium, the partial pressure of water vapour (in Pascals) in the atmosphere $p_{\ce{H2O}}$ and the water mass fraction in the magma ocean $F_{\ce{H2O}}$ are related by \citep{Papale1997}:
\begin{equation}
	p_{\ce{H2O}} = \left( \frac{ F_{\ce{H2O}} }{ 3.44 \times 10^{-8} } \right)^{1/0.74}.
	\label{water_equi}
\end{equation}
An overview of the most important parameters of the volatile model is provided in Table~\ref{Tab_Volat_Model}. 
The values used in our simulations are provided in the appendix in Table~\ref{Tab_Therm_Model_Value}. 

The mass balance of the water contained in the magma ocean and atmosphere system is given by the following equation:
\begin{equation}
	\begin{aligned}
		M_{\ce{H2O}}^{\mathrm{moa}} &= M_{\ce{H2O}}^{\mathrm{crystal}} + M_{\ce{H2O}}^{\mathrm{liq}} + M_{\ce{H2O}}^{\mathrm{atm}} \\
		&= k_{\ce{H2O}} F_{\ce{H2O}} M^{\mathrm{crystal}} + F_{\ce{H2O}} M^{\mathrm{liq}} + \frac{4 \pi r_\mathrm{p}^2}{g} p_{\ce{H2O}}.
	\end{aligned}
\end{equation}

As already mentioned, there are two sinks of water in the system: the solidification of the magma ocean which traps water in the solid mantle and the photolyzation of water followed by atmospheric escape of hydrogen and oxygen at the top of the atmosphere.
Therefore, the water mass in the different reservoirs is governed by two differential equations, taking into account both sinks:
\begin{eqnarray}
	\label{Deriv_Watersol}
	\frac{dM_{\ce{H2O}}^{\mathrm{sol}}}{dt} &=& 4 \pi \rho_\mathrm{m}  k_{\ce{H2O}} F_{\ce{H2O}} r_\mathrm{s}^2 \frac{d r_\mathrm{s}}{dt}, \\
	\frac{dM_{\ce{H2O}}^{\mathrm{moa}}}{dt} &=& - \frac{dM_{\ce{H2O}}^{\mathrm{sol}}}{dt} - 4 \pi r_\mathrm{p}^2 \phi_1 \frac{\mu_{\ce{H2O}}}{2 \mu_{\ce{H}}}.
	\label{Dervi_Watermo}
\end{eqnarray}
We assume a constant, mantle-averaged partition coefficient $k_{\ce{H2O}}$ for water between the melt and solid.
Even though the value of this parameter should vary for different materials, for the purpose of this simplified model a constant coefficient is sufficient.

Due to the XUV flux from the star, the water is dissociated and the hydrogen escapes at the rate $\phi_1$.
The remaining oxygen goes into the magma ocean and oxidizes \ce{FeO} to \ce{Fe2O3}.
The oxygen can be dragged along with the hydrogen and can escape at a rate $\phi_2$ in the case of very high hydrogen loss rate\footnote{See \citet{Hunten1987} for the calculation of the `cross over mass' $m_c$, which determines if a heavy species can be dragged along by a lighter species during atmospheric escape.}, which is likely to happen for planets in close orbits around early M dwarfs due to the particularly high XUV flux emitted by these stars. The escape rates of hydrogen and oxygen, 
$\phi_1$ and $\phi_2$, respectively, are calculated by the \vplanet{} module \atmesc{}, which is based on the atmospheric escape model of \citet{Luger2015a}.

The amount of oxygen in the different reservoirs is governed by a similar set of differential equations:
\begin{eqnarray}
	\label{Deriv_Oxysol}
	\frac{dM_{\ce{O}}^{\mathrm{sol}}}{dt} &=& 4 \pi \rho_\mathrm{m}  F_{\ce{FeO_{1.5}}} r_\mathrm{s}^2 \frac{d r_\mathrm{s}}{dt} \frac{\mu_{\ce{O}}}{2 \mu_{\ce{FeO_{1.5}}}}\\
	\frac{dM_{\ce{O}}^{\mathrm{moa}}}{dt} &=& 4 \pi r_\mathrm{p}^2 \left(  \phi_1 \frac{\mu_{\ce{O}}}{2 \mu_{\ce{H}}} - \phi_2 \right) - \frac{dM_{\ce{O}}^{\mathrm{sol}}}{dt}
	\label{Deriv_Oxymo}
\end{eqnarray}
However, unlike water which dissolves into the magma ocean following Eq. (\ref{water_equi}), the oxygen reacts with the \ce{FeO} in the melt, oxidizing it to \ce{Fe2O3}.

The composition used by \citet{Schaefer2016} and in \magmoc{} is the Bulk Silicate Earth \citep{ONeill1998}.
To simplify our model, we assume that no oxygen builds up in the atmosphere as long as the reservoir of \ce{FeO} in the magma ocean is not fully oxidized.
When there is no \ce{FeO} left in the magma ocean, no more oxygen dissolves into the melt but it builds up in the atmosphere. 

\section{Validation}
\label{chap_validation}

In the following, we present \magmoc{} simulations for two scenarios to validate our model: one set-up for the Super-Earth GJ~1132b and one for Earth and compare to previous work by \citet{Schaefer2016}, \citet{Elkins-Tanton2008}, and \citet{Hamano2013}, respectively.

\subsection{GJ~1132b}
\label{sec_res_GJ}

\begin{figure*}[ht]
    \centering
    \includegraphics[width=0.95\textwidth]{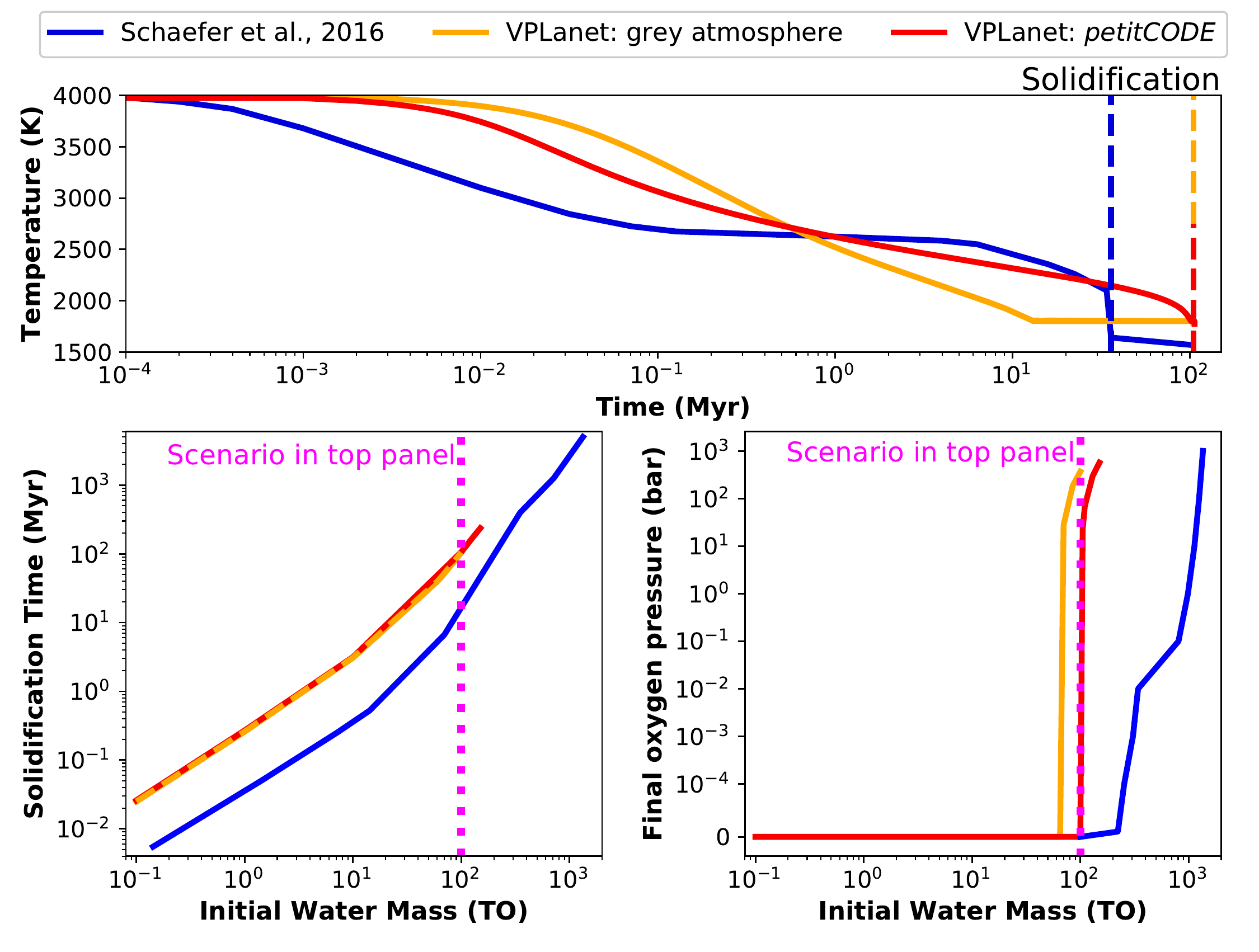}
    \caption{\textit{Upper:} Evolution of the potential mantle temperature of GJ~1132b for $\SI{100}{TO}$ initial water content from \citet[Fig. 4]{Schaefer2016} (blue) compared to the results obtained with \vplanet{} with the grey atmosphere model (yellow) and the \petit{} (red). Solidification times (= atmospheric desiccation) for all three simulations are indicated by dashed lines. Solidification proceeds similarly but slightly faster with \vplanet{} while desiccation of atmosphere occurs later. This scenario is indicated in the lower panels (pink line). \textit{Lower left:} Comparison of the solidification time of the magma ocean for range of initial water masses with different models. Data from \citet[Fig. 5]{Schaefer2016}. \textit{Lower right:} Comparison of final oxygen pressure in the atmosphere at magma ocean solidification. Data from \citet[Fig. 7]{Schaefer2016}.
    The necessary input files to reproduce the results in this paper can be found on our GitHub repository linked in the figure captions:
    \textcolor{purple}{$\nearrow$
    \href{https://github.com/pbfeu/Trappist1_MagmOc/tree/master/Fig_Temp_GJ1132b}{GitHub}}
    }
    \label{GJ1132b_VPLanet_Schaefer}
\end{figure*}

\begin{table}[ht]
	\caption[Parameters GJ~1132b]{Physical parameters of GJ~1132b \citep{Bonfils2018}}
	\begin{tabular}{ccc}
		\noalign{\smallskip}
		\hline
		\noalign{\smallskip}
		Symbol & Parameter & Value \\ 
		\noalign{\smallskip}
		\hline \hline
		\noalign{\smallskip}
		$r_\mathrm{p} $ & Planetary radius & $1.15 \, R_\Earth$ \\
		$M_\mathrm{p} $ & Planetary mass  & $1.62 \, M_\Earth$ \\
		$r_\mathrm{c}$ $^a$ & Core radius & $1.15 \, r_{\mathrm{c},\Earth}$ \\
		$A$ & Albedo (steam atm.)& 0.75 \\
		\noalign{\smallskip}
		\hline
	\end{tabular}
	\\
	$^a$ assumption, $r_{\mathrm{c},\Earth} \sim \SI{3400}{\kilo\metre}$
	\label{Tab_Param_GJ1132b}
\end{table}

GJ~1132b is a Super-Earth, with a size of 1.2 Earth radii orbiting an M dwarf at a distance to Earth of about $\SI{12}{\parsec}$.
With an equilibrium temperature of $\SI{409}{\kelvin}$, assuming an albedo of 0.75, it is too hot to sustain liquid water on the surface and be habitable.
Still, it is an interesting target for observing atmospheric composition and dynamics on a rocky exoplanet \citep{Berta-Thompson2015}.
The physical parameters for GJ~1132b used for this work can be found in the Appendix in Table~\ref{Tab_Param_GJ1132b}.

In order to test \magmoc{}, we compare the results of \vplanet{} for GJ~1132b to those from \citet{Schaefer2016}, in the following called \textit{Schaefer model}. We calculated the atmosphere and magma ocean evolution for 0.1 - 200 terrestrial oceans and compared the results using \petit{} \citep{Molliere2015, Molliere2017} and the grey model \citep{Elkins-Tanton2008, Carone2014} for the atmosphere cooling part with those obtained by \citet{Schaefer2016}.

Figure~\ref{GJ1132b_VPLanet_Schaefer} shows as an example of the resulting cooling curves of the mantle temperature for $\SI{100}{TO}$ initial water content from the Schaefer model and \vplanet{} with both atmospheric models: the grey atmosphere and the \petit{}. For most of the evolution, the mantle cools more slowly in the Schaefer model but it solidifies earlier compared to our model when the atmosphere is desiccated. Only for the first Myr does the Schaefer model cools faster. Note, however, the logarithmic time axis which skews the perception towards the very early evolution.

The reason for the different cooling behavior is different implementations of the climate models: the Schaefer model uses a line-by-line radiative climate model with a simplified temperature-atmosphere profile at the base that follows a dry adiabat from the surface (set by the magma ocean temperature) to the tropopause. We use a similar temperature structure but a simplified radiative treatment in our grey atmosphere model. The \petit{} uses the correlated-k method \citep{Molliere2015, Molliere2017} for radiative transfer and iterates the atmospheric pressure temperature profile self-consistently until emission and absorption are in balance for each vertical atmosphere layer in the radiative region.
Below that, in the convective region, the atmosphere follows a moist adiabatic profile.
The \petit{} produces a cooling curve that matches more closely the results from the Schaefer model than the grey atmosphere does, especially after the first million years. But it leads to a later solidification of the magma ocean like the grey atmosphere model.

In other words, even though the cooling curve is much steeper with the \petit{} and the grey atmosphere model compared to the Schaefer model for large parts of the simulation, i.e. the first $\SI{10}{\mega\year}$, the magma ocean needs more time to solidify.
That is because the time of solidification mainly depends on the atmospheric desiccation, i.e. the loss of all water from the atmosphere towards the end of the magma ocean state. Apparently, the atmosphere in the Schaefer model desiccates more rapidly than both our grey and \petit{} models. A possible reason for this deviation might be a different age of the star at the beginning of the magma ocean simulation. We use a stellar age of $\SI{5}{\mega\year}$ while \citet{Schaefer2016} seem to use no offset.
Because we start the magma ocean evolution later, we receive a lower XUV flux during the evolution and thus need more time to desiccate the atmosphere.
When starting our simulation with a stellar age of 0, the atmosphere is desiccated after $\SI{43}{\mega\year}$ which is only $\sim 15 \%$ larger than the result of \citet{Schaefer2016}.
We argue that the estimated lifetime of the protoplanetary disk \citep[$\sim$5~Myr,][]{Ribas2014} is a more realistic starting time for the simulation of the magma ocean evolution. 
We note that we did not not take the effect of large impactors into account and how these could re-initiate the magma ocean stage.

We also examined the oxygen pressure in the atmosphere at the end of the atmospheric escape for different initial water masses and for the different models (Fig.~\ref{GJ1132b_VPLanet_Schaefer}, \textit{lower right}). Here, we find in our models that the amount of oxygen that builds up in the atmosphere depends crucially on the slope of the cooling curve towards the end of the magma ocean evolution as it fully solidifies.  Due to the faster cooling with the grey atmosphere and the \petit{} compared to the Schaefer model, the \ce{FeO} in the magma ocean becomes fully oxidized much faster for the last few percent of the mantle solidification. Therefore, there is still water escaping from the atmosphere when the oxygen buffer in the mantle is filled and free oxygen starts to build up in the atmosphere.

In the grey atmosphere model, the planet cools down the fastest (see Fig.~\ref{GJ1132b_VPLanet_Schaefer}).
Therefore, oxygen starts to build up already for initial water masses of $\SI{70}{TO}$. 
With the \petit{}, only for initially more than $\SI{100}{TO}$ of water in the system, oxygen builds up in the atmosphere, while the Schaefer model does not start to build up oxygen for initial water masses smaller than $\SI{200}{TO}$.

We did not run simulations for initial water masses larger than $\SI{200}{TO}$ with the \petit{} and $\SI{100}{TO}$ with the grey atmosphere. For one of the most extreme cases we considered ($\SI{100}{TO}$), we find that the steep cooling curve of the mantle also drives catastrophic outgassing that leads to a mass fraction of water in the melt of nearly 20\%. Since even 20\% is a significant fraction of the mantle and our model does not take into account changes in the density or the planet's radius, we did not run simulations with more than $\SI{100}{TO}$ with \vplanet{}/grey and $\SI{200}{TO}$ with \vplanet{}/\petit{}. For $\SI{1000}{TO}$ initial water mass, the mass fraction of water in the melt would reach even 100\%, which we consider as unphysical in this set-up. To consider such high values, we would need to account for changes in density in the equations of state, which are currently not implemented.
When cooling is slower, as is the case in the Schaefer model, the atmospheric escape can balance the outgassing better and the water fraction in the melt stays at a smaller level, even for very high water contents. Therefore, it was possible for \citet{Schaefer2016} to run simulations up to initial water masses of $\SI{1000}{TO}$.

We conclude that the grey atmosphere model based on \citet{Elkins-Tanton2008} and \citet{Carone2014} was successfully adapted and benchmarked with the correlated-k \petit{} \citep{Molliere2015, Molliere2017} to be applicable to a primordial steam atmosphere on a magma ocean world. 
Generally, we reproduce the results of \citet{Schaefer2016} with \magmoc{} using both the \petit{} and the grey atmosphere model (Fig.~\ref{GJ1132b_VPLanet_Schaefer}): 
\begin{itemize}
    \item Larger initial water masses lead to longer solidification times, where our solidification times for both atmosphere models are consistently larger by a little less than one order of magnitude compared to \citet{Schaefer2016},  
    \item Abiotic oxygen builds up for about $\SI{100}{TO}$ initial water mass. The exact initial water mass limit, for which onset of oxygen build-up happens, varies within half an order of magnitude (between $\SI{80}{TO}$ for the grey atmosphere model to $\SI{200}{TO}$ for \citet{Schaefer2016}).
\end{itemize}
We will show henceforth \magmoc{} results for the TRAPPIST\=/1 planets and Earth using the grey atmosphere model.

\subsection{Earth}
\label{sec: Earth}

\begin{table}[ht]
    \caption{\vplanet{} input parameters for Earth}
	\begin{tabular}{cc}
		\noalign{\smallskip}
		\hline
		\noalign{\smallskip}
		Parameter & Initial value\\ 
		\noalign{\smallskip}
		\hline \hline
		\noalign{\smallskip}
		$M_{\ce{H2O}}^\mathrm{ini} $ & $2-\SI{20}{TO}$ \\
		$T_\mathrm{surf}^\mathrm{ini} = T_\mathrm{p}^\mathrm{ini} $ & $\SI{4000}{\kelvin}$ \\
		$\epsilon_\mathrm{XUV}$ & 0.3  \\ 
		Atmosphere model & grey \\
		\vplanet{} modules used & \magmoc{}, \atmesc{}, \radheat{}, \stellar{} \\
		\noalign{\smallskip}
		\hline
	\end{tabular}
	\label{Tab_Input_Earth}
\end{table}

\begin{table*}[ht]
\begin{center}
    \caption{\vplanet{} Results for Earth: Comparison between \citet[Tab. 3, Earth ($\SI{2000}{\kilo\metre}$)]{Elkins-Tanton2008} and \vplanet{} (without \ce{CO2}) \textcolor{purple}{$\nearrow$
    \href{https://github.com/pbfeu/Trappist1_MagmOc/tree/master/Tab_Earth_Elkins-Tanton}{GitHub}}}
	\begin{tabular}{c|cc|cc}
		\noalign{\smallskip}
		\hline
		\noalign{\smallskip}
		Initial volatile mass & \multicolumn{2}{c|}{Dry case: $\SI{2}{TO}$ \ce{H2O} / $\SI{0.4}{TO}$ \ce{CO2}} & \multicolumn{2}{c}{Wet case: $\SI{20}{TO}$ \ce{H2O} / $\SI{4}{TO}$ \ce{CO2}} \\ 
		\noalign{\smallskip}
		\hline
		\noalign{\smallskip}
		Model & Elkins-Tanton & \vplanet{} (w/o \ce{CO2}) & Elkins-Tanton & \vplanet{} (w/o \ce{CO2})  \\
		\noalign{\smallskip}
		\hline \hline
		\noalign{\smallskip}
		\multicolumn{5}{l}{Fraction of initial water content degassed into atmosphere [\%]} \\
		\noalign{\smallskip}
		\hline
		\noalign{\smallskip}
		& 70 & 94 & 91 & 96 \\
		\noalign{\smallskip}
		\hline
		\noalign{\smallskip}
		\multicolumn{5}{l}{Final atmospheric pressure (sum of partial pressures of \ce{H2O} and \ce{CO2}) [bar]} \\
		\noalign{\smallskip}
		\hline
		\noalign{\smallskip}
		& 240 & 499 & 3150 & 5083 \\
		\noalign{\smallskip}
		\hline
		\noalign{\smallskip}
		\multicolumn{5}{l}{Time to reach 98\% solidification, for $k_{\ce{H2O}} = 0.01$ and $k_{\ce{CO2}} = 0.001$ [Myr]} \\
		\noalign{\smallskip}
		\hline
		\noalign{\smallskip}
		& 0.06 & 0.3 & 2.4 & 3.0 \\
		\noalign{\smallskip}
		\hline
		\noalign{\smallskip}
		\multicolumn{5}{l}{Water content of liquids remaining at 98\% solidification [$\si{\wtpercent]}$} \\
		\noalign{\smallskip}
		\hline
		\noalign{\smallskip}
		& 1.5 & 1.7 & 5.3 & 9.5 \\
		\noalign{\smallskip}
		\hline
	\end{tabular}
	\label{Tab_Results_Earth}
\end{center}
\end{table*}

We also validated our magma ocean\=/atmosphere model for the case of the young Earth. Similar to other terrestrial planets, Earth is believed to have had a deep magma ocean due to all the energy deposited during accretion \citep[\eg][]{Elkins-Tanton2008,Lammer2018A}. During the solidification of this magma ocean, some of the interior water outgassed into the atmosphere.
Unlike GJ~1132b, which is orbiting an M dwarf, Earth is orbiting a G star at a much larger distance and the Sun is not as active as an M dwarf.
Therefore, the atmospheric escape on early Earth was much weaker than on the young GJ~1132b.

Earth's initial water content is weakly constrained because even the Earth's current total water content is uncertain. In addition to the one Earth ocean of water located on the Earth's surface, much more water might be locked for example in the Earth's mantle transition zone at a depth of $410 - \SI{660}{\kilo\metre}$ \citep{Pearson2014, Schmandt1265}. 

\begin{figure}[h]
    \centering
    \includegraphics[width=\columnwidth]{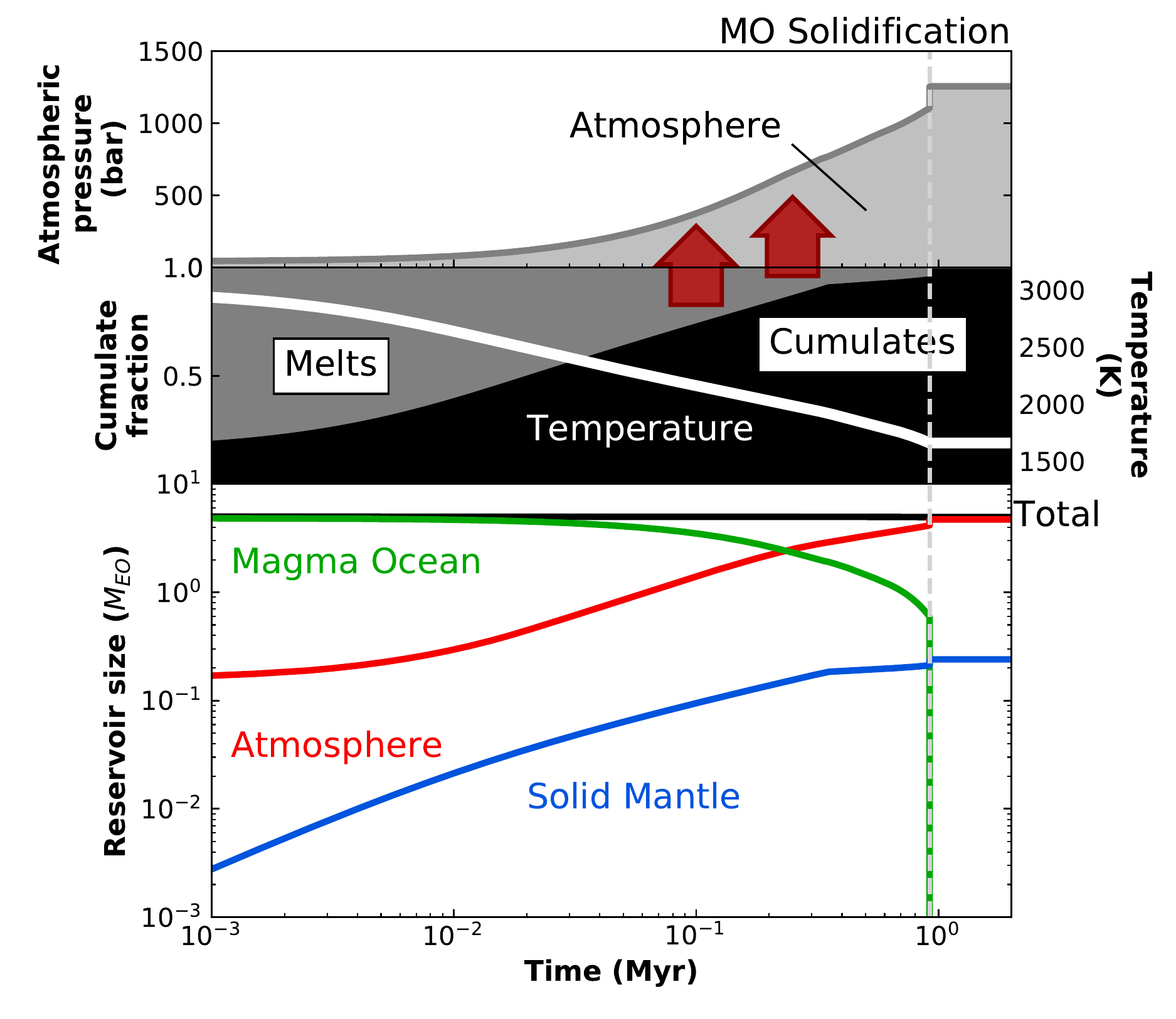}
    \caption{Simulations of magma ocean evolution on Earth with \vplanet{}/\magmoc{} for $\SI{5}{TO}$ initial water content. \textit{Top to bottom:} Atmospheric pressure of water. Surface temperature and solid fraction of the mantle. Water masses in the different reservoirs. The red arrows indicate the catastrophic outgassing into the atmosphere due to the decreasing of the melt reservoir. The results are very similar to simulations presented by \citet[Fig. 1]{Hamano2013}, especially the final water masses and the atmospheric pressure. With \magmoc{}, the magma ocean solidifies after $\SI{0.9}{\mega\year}$ and with \citet{Hamano2013} after $\SI{4}{\mega\year}$. That is, the solidification times differ by a factor of roughly 4.
    \textcolor{purple}{$\nearrow$
    \href{https://github.com/pbfeu/Trappist1_MagmOc/tree/master/Fig_Earth_Hamano}{GitHub}}}
    \label{Earth_Hamano}
\end{figure}

\citet{Hamano2013} present simulations of terrestrial planets, including Earth, with a water atmosphere containing $\SI{5}{TO}$ and in a magma ocean phase:
Their results show that Earth solidifies rapidly, after $\sim \SI{4}{\mega\year}$ for an initial water mass of $\SI{5}{TO}$ ($T_\mathrm{ini} = \SI{3000}{K}$, $A = 0.3$). Figure~\ref{Earth_Hamano} shows the results of simulations with \magmoc{} for Earth with the same initial water content ($T_\mathrm{ini} = \SI{3000}{K}$, $A = 0.75$). The red arrows indicate how the solidification of the magma ocean, i.e. the decreasing size of the melt reservoir, leads to the very strong or catastrophic outgassing of water into the atmosphere \citep{Elkins-Tanton2008,Lammer2018A}. 

Even though we use a different model for the outgoing flux, which leads to a different shape of the cooling profile, the solidification time differ only by a factor of 4 (\magmoc: $t_\mathrm{sol} \sim \SI{0.9}{\mega\year}$, \citet{Hamano2013}: $t_\mathrm{sol} \sim \SI{4}{\mega\year}$). Also, the resulting water masses in the different reservoirs are the same for both simulations (See Fig.~\ref{Earth_Hamano} and \citet[Fig. 1]{Hamano2013}).

For GJ~1132b, the magma ocean is initially prevented from effective cooling by the extreme greenhouse effect of the water vapor atmosphere, since the planet is so close to its star. It can, however, cool off after atmospheric escape of hydrogen has reduced the atmospheric mass. Thus, atmospheric escape is in this case the driving factor for magma ocean solidification. In our GJ~1132b simulations, we tend to observe similar but slightly later (within one order of magnitude) solidification times compared to \citet{Schaefer2016}. Since Earth is farther away from its host star than GJ~1132b, the solidification time is mainly determined by the implementation of the cooling flux. In our Earth simulation, this leads to a similar but slightly earlier (by a factor of 3) solidification time compared to \citet{Hamano2013}.

\citet{Elkins-Tanton2008} presents simulations of the evolution of a magma ocean on Earth for different amounts of water and \ce{CO2} in the atmosphere. For our simulations of the early Earth, we used the values as listed in Table~\ref{Tab_Input_Earth} that are based on \citet{Elkins-Tanton2008}. We also chose a magma ocean depth of 2000~km to be consistent with \citet{Elkins-Tanton2008}.
However, we did not include the effect of \ce{CO2} in our model.
We compare our results with a pure water atmosphere to her results in Table \ref{Tab_Results_Earth}.

Especially for the water rich case ($\SI{20}{TO}$ water and $\SI{4}{TO}$ \ce{CO2}), the results are very similar and differ by less than a factor of two compared to our Earth simulation with a pure water atmosphere simulation. We derive, however, consistently larger final atmosphere pressures and solidification time. We attribute these differences mainly to a different prescription of equations of states for \ce{H2O}, because we adopted the density profiles used by \citet{Schaefer2016}, which differ from those used by \citet{Elkins-Tanton2008}.

The results presented in this section are qualitatively - within one order of magnitude - similar to \cite{Schaefer2016}, \cite{Elkins-Tanton2008}, and \citet{Hamano2013}.
This shows that our model is able to simulate the magma ocean evolution of different kinds of rocky planets: in- and outside of their star's habitable zone.
The quantitative differences are within observational uncertainties. We therefore conclude that our model successfully reproduces past results and is validated. 

We further note here, that for both, GJ~1132b and also for Earth, we find that only $3-5\%$ of the initial water will be locked in the mantle after the magma ocean solidified. Thus, it appears that for Earth, the intermediately wet formation scenario regime (10 - \SI{100}{TO}) is favored to derive the current low water content of 3-\SI{10}{TO}. This conclusion, however, comes with the caveat that we did not consider the influence of the Moon impactor and other smaller impactors. E.g. \citet{zahnle2019} recently showed that smaller impactors can substantially change mantle and atmosphere chemistry and remove water from the system.
Also, the moon forming impact was only the last of multiple events leading to large-scale mantle melting \citep{Jacobson2014,Bottke2015}.
Two episodes of magma oceans are needed to explain the isotopic difference of \ce{^3He}/\ce{^{22}Ne} between the depleted mantle and a primitive deep reservoir.
The D/H ratio of sea water indicates that the silicate Earth during the last global magma ocean was oxidized enough to outgas \ce{H2O} and \ce{CO2} and not \ce{H2} and \ce{CO} \citep{Pahlevan2019}.
The latter, reducing atmosphere mixture would heave facilitated atmospheric escape of hydrogen leading to a higher D/H ratio.
Similarly, the D/H ratio implies a short-lived steam atmosphere, i.e. a short solidification time of the magma ocean \citep{Stueken2020}.
Both circumstances, the small effect of atmospheric escape of hydrogen and the short solidification times, are represented in our model.
We now turn to applying \magmoc{} to the potentially habitable planets of the TRAPPIST-1 system. 

\section{Results for the TRAPPIST-1 planetary system}
\label{chap_results}

\begin{table*}[ht]
\begin{center}
    \caption{Physical and run parameters for TRAPPIST\=/1~e, f, and g.}
	\begin{tabular}{cccc}
		\noalign{\smallskip}
		\hline
		\noalign{\smallskip}
		Planet & TRAPPIST\=/1 e  & TRAPPIST\=/1 f & TRAPPIST\=/1 g \\ 
		\noalign{\smallskip}
		\hline \hline
		\noalign{\smallskip}
		$r_\mathrm{p} \: [R_\Earth]$ ${}^{a}$ &0.920 & 1.045 & 1.129  \\
		$r_\mathrm{c} \: [R_\Earth]$ ${}^{b}$ &0.490 & 0.557 & 0.602  \\
		$M_\mathrm{p} \: [M_\Earth]$ ${}^{a}$  & 0.692 & 1.039 & 1.321 \\
		$a \: [\si{\astronomicalunit}]$ ${}^{a}$ & 0.0293 & 0.0385 & 0.0468 \\
		$e$ ${}^{c}$ & 0.005 & 0.01 & 0.002 \\
		\noalign{\smallskip}
		\hline
		\noalign{\smallskip}
		escape stop time [Myr] & 253.2 & 129.4 & 76.3 \\
		\noalign{\smallskip}
		\hline
		\noalign{\smallskip}
		Albedo &  \multicolumn{3}{c}{0.75} \\
		$M_{\ce{H2O}}^\mathrm{ini}$ &  \multicolumn{3}{c}{$1-\SI{100}{TO}$} \\
		$T_\mathrm{surf}^\mathrm{ini} = T_\mathrm{p}^\mathrm{ini}$  & \multicolumn{3}{c}{\SI{4000}{\kelvin}} \\
		$\epsilon_\mathrm{XUV}$ &  \multicolumn{3}{c}{0.3}  \\ 
		Stellar age at $t=0$ & \multicolumn{3}{c}{\SI{5}{\mega\year}}  \\ 
		Atmospheric model &  \multicolumn{3}{c}{grey} \\
		\vplanet{} modules used & \multicolumn{3}{c}{\magmoc{}, \atmesc{}, \radheat{}, \eqtide{}, \stellar{}} \\
		\noalign{\smallskip}
		\hline
		\noalign{\smallskip}
	    \multicolumn{4}{l}{Heating power (TW) for reference/extreme heating at beginning of simulation ($t=0$):} \\
		\noalign{\smallskip}
		\hline
		\noalign{\smallskip}
		Radiogenic heating & 57 / $2.8 \times 10^4$ & 69 / $3.3 \times 10^4$ & 85 / $4.1 \times 10^4$\\
		Tidal heating & 3.9 / $1.6 \times 10^3$ & 4.1 / 49 & 0.06 / 18\\
		\noalign{\smallskip}
		\hline
		\noalign{\smallskip}
	    \multicolumn{4}{l}{Current water fraction estimates [wt\%] (values in brackets show min and max water mass fraction):} \\
		\noalign{\smallskip}
		\hline
		\noalign{\smallskip}
		\citet{Dorn2018} ${}^{d}$ & $2^{+2}_{-1}$ & $7^{+3}_{-3}$ & $13^{+4}_{-4}$ \\
		\citet{barr2018interior} ${}^{e}$ & 0-98 & 32-99 & 32-60\\
		\citet{Unterborn2018b} ${}^{d}$ & 0-1 & 2-11 & 9-23 \\
		\citet{Agol2020} ${}^{f}$ & $0.3^{+1.8}_{-0.3}$ & $1.9^{+1.5}_{-1.3}$ & $3.5^{+1.6}_{-1.3}$ \\
		This study, $M_\mathrm{p}/r_\mathrm{p}$ 1$\sigma$ ${}^{g}$
		& 11.0$\pm$10.2 (0-23) & 7.9$\pm$8.1 (1-21) & 18.4$\pm$5.6 (11-24) \\
		\noalign{\smallskip}
		\hline
	\end{tabular}
	\\
	${}^{a}$ \citet{Agol2020}, ${}^{b} r_\mathrm{c} = r_\mathrm{p} \times r_{\mathrm{c,\Earth}}/R_\Earth$, ${}^{c}$ \citet{Grimm2018}, ${}^{d}$ based on \citet{Grimm2018} data, ${}^{e}$ based on \citet{Gillon2017} data, ${}^{f}$ for a core mass fraction of 25\%, ${}^{g}$ see Fig.~\ref{fig:InStr_all}, model by \citet{Noack2016}, based on \citet{Agol2020} data
	\label{Tab_Input_TRAPPIST-1}
\end{center}
\end{table*}

\begin{figure*}[ht]
    \centering
    \includegraphics[width=2\columnwidth]{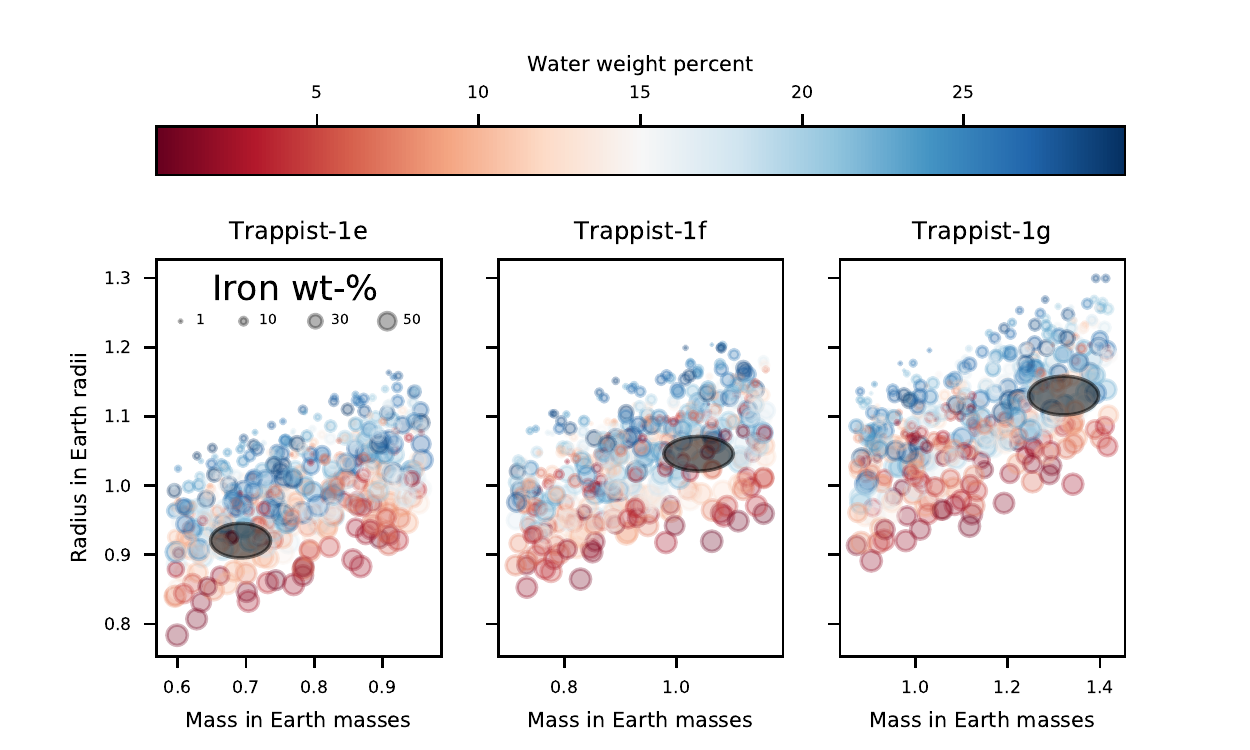}
    \caption{Mass-radius data for 1000 interior structure models per planet using randomly-selected compositions (in terms of water and iron fraction) and main planetary parameters. The grey error circles give the ranges of measured masses and radii from \citet{Agol2020}.}
    \label{fig:InStr_all}
\end{figure*}

Since TRAPPIST\=/1~e, f, and g are orbiting TRAPPIST\=/1 in its habitable zone \citep[\eg][]{Kasting1993,Abe2011,Kopparapu2014,Catling2018,Turbet2018,Chen2019}, we concentrated our efforts on simulating the magma ocean and volatile evolution for these three planets. Studies by \citet{Agol2020} and \citet{Grimm2018} provided new constraints on the radii and masses of the TRAPPIST\=/1 planets (Table~\ref{Tab_Input_TRAPPIST-1}). With these constraints, it is possible to give estimates on the density and composition of the planets \citep{Dorn2018,Unterborn2018b,barr2018interior}. 

To have a better understanding of the range of water fractions in TRAPPIST\=/1~e, f, and g, depending on several random parameters such as the interior and surface temperatures, we used for comparison our interior structure model \citep{Noack2016}. With this model, we investigated the range of compositions for TRAPPIST\=/1~e, f, and g based on the masses and radii as listed in \citet{Agol2020} for a 1$\sigma$ error range. 
Our interior structure model solves composition-dependent equations of state for iron, different rock assemblages, high-pressure ices and liquid water. The transition from liquid water to ice as well as the correct high-pressure ice phase (ice IV, VI, VII, VIII or X) is determined self-consistently for an adiabatic temperature profile. We assume arbitrary compositions for the three planets, where the iron weight fraction and the water fraction are randomly selected between 0 and 70 wt\%. We also account for different iron distributions between mantle and core, where we allow between 0 and 20\% of the mantle minerals to be iron-based. Other parameters, such as mass (within the known error range), surface pressure (between 1 and $\SI{100}{\bar}$), interior temperature profile and the surface temperature (effective temperature plus up to 50 degrees to account for greenhouse gases) are also selected randomly.

Figure \ref{fig:InStr_all} shows the entire range of radii that are predicted for TRAPPIST-1 e, f and g by our model plotted over the input mass. The colour coding indicates the water content, and the size of the dots show the planet iron fraction. Grey circles indicate the range of masses and radii from the estimates by \citet{Agol2020} reported in Table \ref{Tab_Input_TRAPPIST-1}. 
The average predicted water concentration together with a 1$\sigma$ standard deviation and the total interval of observed water fractions is given in Table \ref{Tab_Input_TRAPPIST-1}.

This evaluation as well as the comparison to published data reported in Table \ref{Tab_Input_TRAPPIST-1} show, that the uncertainty in the current water content remains
large for the TRAPPIST\=/1 planets. We thus simulate the magma ocean and volatile evolution of the planets TRAPPIST\=/1~e, f, and g for initial water masses from $1-\SI{100}{TO}$ until the planets become desiccated or until atmospheric escape stops.

In this work, we simulate the dominant process that leads to significant atmospheric erosion during the pre-main sequence of the host star, which is contained in the \vplanet{} module \atmesc{}, described in more detail in \citet{Barnes2020}. 
More precisely, \atmesc{} uses energy- and diffusion limited hydrogen and oxygen escape following \citet{Luger2015a,Luger2015}. 
Hydrogen is produced when the absorbed XUV flux of the host star photodissociates water, where we use an XUV absorption efficiency of $\epsilon_\mathrm{XUV} = 0.3$\footnote{We choose this value to be consistent with \citet{Schaefer2016}. However, we note that other models of water loss assume lower values for $\epsilon_\mathrm{XUV}$ \citep{Bolmont2016,Lincowski2018}, so we may be overestimating water loss.} (see also Table~\ref{Tab_Input_TRAPPIST-1}). 
For the TRAPPIST\=/1 system, we use the stellar evolution model contained in the \vplanet{} module \stellar{}. It assumes the constant-followed-by-power-law model of \citet{Ribas2005} for XUV-flux and calculates the total bolometric luminosity from grids based on \citet{Baraffe2015}. The stellar luminosity of TRAPPIST\=/1 decreases with time. At some point during the evolution, typically after the magma ocean solidification, the incoming flux at TRAPPIST\=/1 e, f, and g  reaches the threshold $S_\mathrm{eff}$, where we use equation~(2) of \citet{Kopparapu2013} and choose their values for the runaway greenhouse limit. This limit depends on the mass of the planet \citep{Kopparapu2014}.
At this  point, we assume that the pure steam atmosphere condenses out. 
Furthermore, the stratosphere desiccates and atmospheric escape of \ce{H2O} stops because water vapour no longer reaches the upper atmosphere to be photolysed.

To set up the model for the TRAPPIST-1 simulations, we choose the following specific assumptions as input for the equations described in Sections~\ref{sec_therm_model}.
For the abundance of radioactive isotopes we use Earth abundances, scaled by the mass of the planet. 
In addition, we consider tidal heating due to eccentric orbits with fixed eccentricities, as shown in Table~\ref{Tab_Input_TRAPPIST-1}. 
We did run simulations with different initial rotation periods and found that rotational tidal heating does not effect the long term evolution of the magma ocean as the rotation decays over a very short time scale.
To decrease computational requirements, we therefore decided to start our simulations with the planets tidally locked and use a tide model with a constant phase lag (CPL) and set $Q$ to 100.
A concise compilation of the input parameters for the TRAPPIST\=/1 planets and the settings for \vplanet{} can be found in Table~\ref{Tab_Input_TRAPPIST-1}. 

\subsection{Different Evolution Scenarios}

\begin{figure*}[ht]
    \centering
    \includegraphics[width=\textwidth]{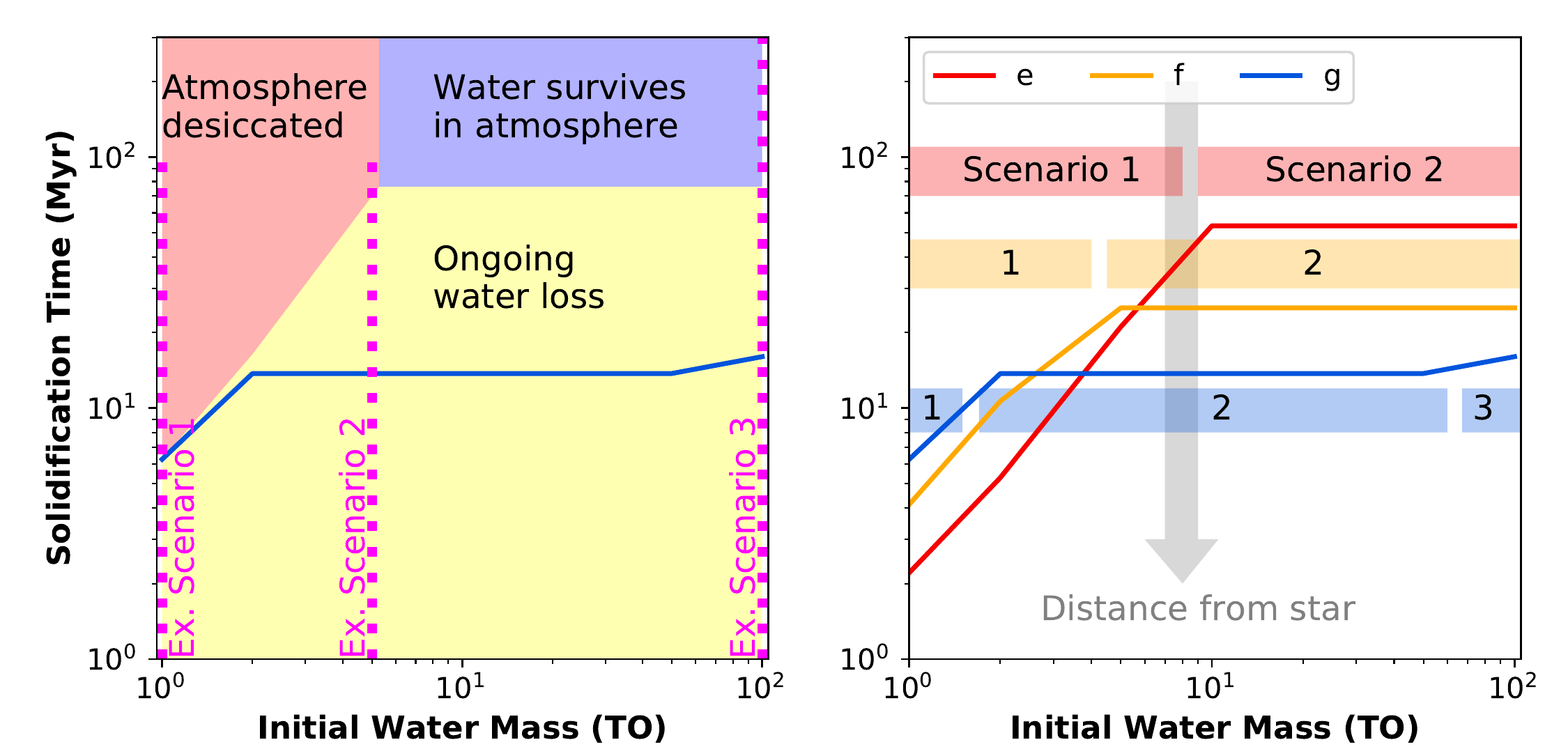}
    \caption{\textit{Left:} Solidification time of the mantle for TRAPPIST\=/1~g. \textit{Yellow:} Ongoing water loss through atmospheric escape; \textit{Red:} Scenarios with initial water masses of less than $\SI{7}{TO}$ will lead to a desiccated atmosphere (no water left); \textit{Blue:} Scenarios with $>\SI{5}{TO}$ water will be left in atmosphere when the atmospheric escape stops \citep{Kopparapu2013}. Vertical lines indicate three examples of the three different characteristic scenarios (see text and Fig.~\ref{Plot_TR1_scenarios_evolution} for more information). \textit{Right:} Solidification and desiccation times for planets e, f, and g (solid lines in red, yellow and blue, respectively). Range of the different scenarios is indicated with colored bars. The grey arrow indicates the distance to the star which increases from planet e to g. The range of the different scenarios depends on the orbital distance of the planet to the star which determines the incoming stellar radiation and therefore the cooling flux.
    \textcolor{purple}{$\nearrow$
      \href{https://github.com/pbfeu/Trappist1_MagmOc/tree/master/Fig_Trappist1g_scenarios}{GitHub}}}
    \label{Summary_Trappist1_scenarios_g}
\end{figure*}

\begin{figure*}[p]
    \centering
    \includegraphics[width=\textwidth]{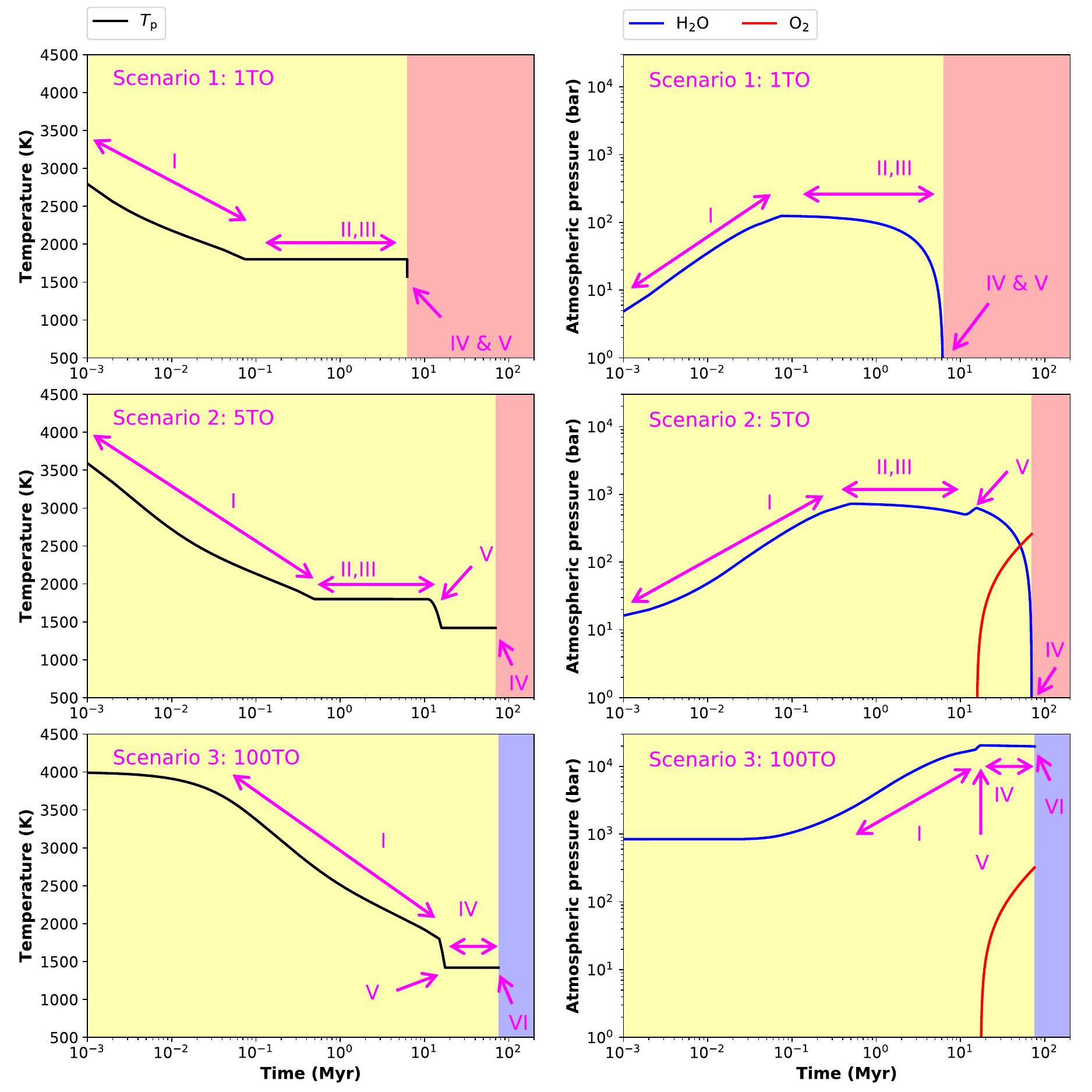}
    \caption{Individual evolution plots for TRAPPIST\=/1~g for the three scenarios indicated in Fig.~\ref{Summary_Trappist1_scenarios_g}: Mantle temperature (\textit{left}) and atmospheric pressure of water and oxygen (\textit{right}). Colors correspond to shaded areas in Fig.~\ref{Summary_Trappist1_scenarios_g} (\textit{Yellow:} Ongoing escape, \textit{Red:} All water lost from atmosphere, \textit{Blue:} Water remains in atmosphere). Numbers indicate important phases in the evolution (see text): (I) Rapid solidification and outgassing, (II) Runaway greenhouse prevents solidification, (III) Outgassing drives escape, (IV) Atmospheric desiccation, (V) Mantle solidification, (VI) \ce{H2O} escape stops.
    \textcolor{purple}{$\nearrow$~\href{https://github.com/pbfeu/Trappist1_MagmOc/tree/master/Fig_Trappist1g_scenarios}{GitHub}}}
    \label{Plot_TR1_scenarios_evolution}%
\end{figure*}

We identify in this work three evolutionary scenarios for the TRAPPIST\=/1 planets, depending on the initial water content and their distance to TRAPPIST\=/1: These are scenario~1 (dry, $<\SI{2}{TO}$ for planet g), 2 (intermediate, $2-\SI{100}{TO}$), 3 (extremely wet, $>\SI{100}{TO}$). The left panel of Fig.~\ref{Summary_Trappist1_scenarios_g} shows the solidification time of the magma ocean for TRAPPIST\=/1~g for initial water masses of $1-\SI{100}{TO}$.
Colored areas show how long the atmospheric escape continues (yellow) and whether the evolution results in a dry (red) or wet (blue) atmosphere.
For each of the three scenarios, representative cases, described in more detail in the next subsections and in Fig.~\ref{Plot_TR1_scenarios_evolution}, are indicated by vertical lines. 
This figure further shows that a TRAPPIST\=/1~g scenario with low water content and short magma ocean solidification times runs the risk of complete desiccation (red region). 
Higher water content and solidification times larger than 14~Myr, on the other hand, would keep water vapour in the atmosphere after the end of the magma ocean state.

In addition, we show the solidification times and input water ranges for the three different scenarios for all three planets in the right panel of Fig.~\ref{Summary_Trappist1_scenarios_g}. 
Note that the border between the scenarios moves to lower water masses with increasing distance to the star.
Figure~\ref{Plot_TR1_scenarios_evolution} shows the evolution of the temperature and the atmospheric pressures for three cases identified in Fig.~\ref{Summary_Trappist1_scenarios_g} (left panel) with numbers indicating important phases of the evolution.
These phases are described in detail in the following for the three representative cases indicated in Fig.~\ref{Summary_Trappist1_scenarios_g} (left):

\subsection*{Example for scenario~1 (dry): Initial water content $\SI{1}{TO}$ for TRAPPIST\=/1~g}

An example of the dry evolution (scenario~1) for TRAPPIST\=/1~g is shown in Fig.~\ref{Plot_TR1_scenarios_evolution}, top panel. The evolution of the temperatures and volatile reservoirs passes through four stages:

\begin{enumerate}
    \item[I] $< \SI{0.1}{\mega\year}$: Rapid onset of solidification and outgassing \\
    The solidification of the magma ocean leads to a rapid increase in the atmospheric pressure as the solid rock cannot store as much water as the liquid melt.
    \item[II] $> \SI{0.1}{\mega\year}$: Runaway Greenhouse \\
    When the surface temperature drops to $\SI{1800}{\kelvin}$, the absorbed stellar radiation is larger than the greenhouse limit of $\SI{280}{\watt\per\square\metre}$ \citep{Goldblatt2013}. Thus, the planet is not able to cool any further.
    \item[III] Atmospheric escape drives outgassing \\
    Since water is escaping at the top of the atmosphere and the magma ocean has not yet completely solidified (the magma melt fraction $\psi$ is still larger than 0.4), more water must outgas into to the atmosphere to maintain the equilibrium between the pressure of water vapour in the atmosphere and the water mass fraction in the magma ocean. However, since atmospheric escape is more efficient than the outgassing from the interior, the pressure in the atmosphere decreases rapidly.
    \item[IV,V] $\sim \SI{6}{\mega\year}$: Atmospheric desiccation \& solidification\\
    Once the atmosphere loses all its water, the planet's cooling rate is no longer restricted by the runaway greenhouse limit and the magma ocean cools rapidly. We stop the evolution when atmospheric water pressure becomes zero, which prevents abiotic oxygen accumulation.
\end{enumerate}

In scenario~1, the planet ends its magma ocean evolution with neither water vapour nor abiotically created \ce{O2} in the atmosphere. We did not consider here the effect of \ce{N2} or \ce{CO2} gas left in the system, which could form the main constituents of the atmosphere in this scenario.

\subsection*{Example for scenario~2 (intermediate): Initial water content $\SI{5}{TO}$ for TRAPPIST\=/1~g}

Here, an example of the intermediate water content case (scenario~2) is shown for TRAPPIST\=/1~g, as shown in the middle panel of Fig.~\ref{Plot_TR1_scenarios_evolution}.

\begin{enumerate}
    \item[I] $< \SI{0.5}{\mega\year}$: Rapid onset of solidification and outgassing \\
    Rapid onset of solidification of the magma ocean leads to outgassing of a thick steam atmosphere.
    \item[II] $> \SI{0.5}{\mega\year}$: Runaway greenhouse \\
    Further, cooling of the magma ocean surface to temperatures below $\SI{1800}{\kelvin}$ is again prohibited by the runaway greenhouse limit.
    \item[II] Atmospheric escape drives outgassing \\
    Escape at the top of the atmosphere drives water out of the magma ocean. Since the atmosphere is very thick compared to scenario~1, atmospheric escape is in this scenario not able to decrease the pressure efficiently as long as the mantle is not yet solidified.
    \item[V] $\sim \SI{14}{\mega\year}$: Mantle solidification\\
    The stellar flux has decreased such that the planetary thermal emission (which is in energy balance with the input stellar irradiation) drops below the greenhouse limit. This allows the magma ocean surface to cool, the temperature to drop below 1800~K and the mantle to solidify. The magma melt fraction $\psi$ becomes smaller than 0.4, marking the end of the magma ocean stage. There is still a thick water atmosphere present at the end of the magma ocean evolution.
    \item[IV] $\sim \SI{50}{\mega\year}$: Atmospheric desiccation \\
    The solidification of the mantle stops the efficient outgassing of water into the atmosphere. Atmospheric escape now removes water from the atmosphere efficiently. At the same time abiotic oxygen builds up.
\end{enumerate}

In scenario~2, the planet ends its magma ocean evolution with a partially or completely eroded steam atmosphere and as a direct consequence with a significant amount of abiotically created \ce{O2} ($>100$~bar). Since the magma ocean is already completely solidified at this point, \ce{O2} can not be efficiently removed from the atmosphere in our model anymore. In the specific case depicted in the middle panels of Fig.~\ref{Plot_TR1_scenarios_evolution}, with a relatively low initial water mass ($\SI{5}{TO}$), the steam atmosphere is completely eroded and only oxygen remains.

\subsection*{Example for scenario~3 (extremely wet): Initial water content $\SI{100}{TO}$ for TRAPPIST\=/1~g}

Here, an example of an evolution track for the very wet evolution (scenario~3) is shown for TRAPPIST\=/1~g, where we depict the magma ocean and surface temperature, water vapour and \ce{O2} atmosphere content in the bottom panel of Fig.~\ref{Plot_TR1_scenarios_evolution}.

\begin{enumerate}
    \item[I] $> \SI{1}{\mega\year}$: Onset of solidification and outgassing \\
    The extremely thick steam atmosphere slows down the solidification rate such that the runaway greenhouse limit (II) is not reached before the end of the simulation in this scenario. Efficient outgassing that increases the atmospheric pressure is also delayed.
    \item[V] $\sim \SI{20}{\mega\year}$: Mantle solidification \\
    The mantle eventually solidifies, slightly later than in scenario~2 because of the slow cooling rate through the thick steam atmosphere.
    \item[IV] Atmospheric escape continues \\
    By now, the stellar XUV flux is very low and thus the atmospheric escape is inefficient. The atmospheric pressure decreases only sightly. At the same time, abiotically created \ce{O2} builds up. However, since this scenario started with a much larger water reservoir than scenario~2, the final oxygen pressure is much lower than the remaining water pressure.
    \item[VI] $\sim \SI{76}{\mega\year}$: \ce{H2O} atmospheric escape stops\\
    The stellar incoming flux at the location of the planet drops below the moist greenhouse threshold. \ce{H2O} condenses out, the stratosphere dessicates and the atmospheric escape of \ce{H2O} stops. A thin magma ocean is still present when the planet enters this regime. However, the viscosity of the remnant magma layer is so high that it cannot efficiently heat the atmosphere anymore. Water condensation at the surface will likely lead to rapid solidification of the remnant magma ocean at this point.
\end{enumerate}

In scenario~3, the planet ends its magma ocean evolution, when an extremely thick steam atmosphere ($>10000$~bar) with about 1\% free \ce{O2} starts to condense out to form an ocean, leaving potentially a 100~bar \ce{O2} atmosphere behind. However, we did not implement processes that could draw down \ce{O2} into the mantle or into a water ocean after solidification.

\subsection{Influence of Additional Heat Sources}

\begin{figure*}[p]
    \centering
    \includegraphics[width=\textwidth]{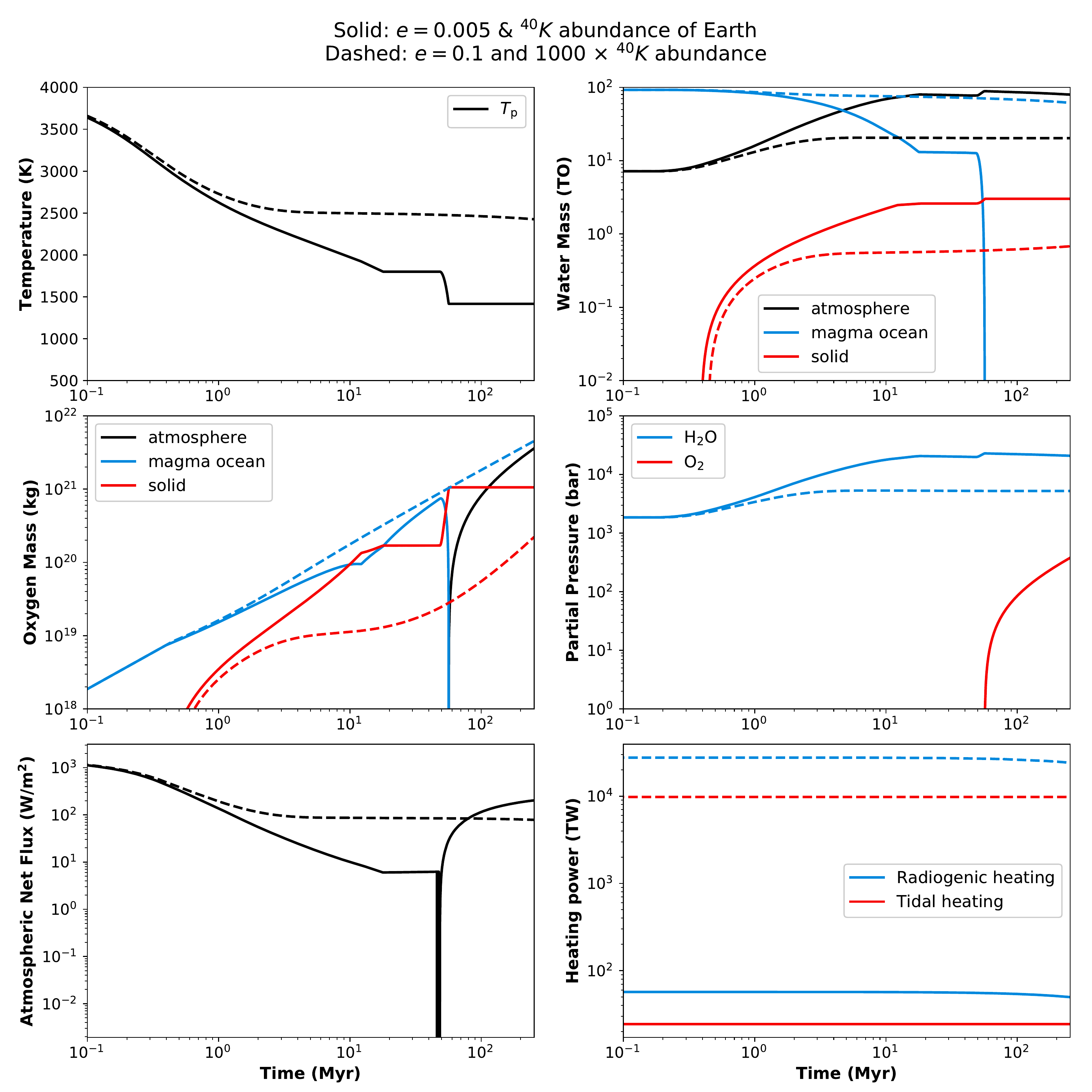}
    \caption{Temperature, volatile budgets (water and \ce{O2}), atmospheric net flux, and heating sources evolution on TRAPPIST\=/1~e with $\SI{100}{TO}$ initial water content for two different heating scenarios: Reference heating (solid), i.e. low eccentricity and Earth's abundance of \ce{^{40}K}, and extreme heating (dashed), i.e. high eccentricity and 1000 times Earth's abundance of \ce{^{40}K}.
    \textcolor{purple}{$\nearrow$\href{https://github.com/pbfeu/Trappist1_MagmOc/tree/master/Fig_Trappist1e_Compare_Heat}{GitHub}}}
    \label{TR1_fluxes_volatiles}%
\end{figure*}

We also tested the influence of additional internal heat sources on the magma ocean evolution of the TRAPPIST\=/1 planets by increasing the radiogenic and tidal heating. 
We found that we need to increase the internal heating enormously to generate substantial changes in the magma ocean evolution. 

We implement this extreme heating by fixing the eccentricity to 0.1, thus mimicking the perturbation so of other planets, and by setting the abundance of \ce{^{40}K} to 1000 times the abundance of Earth. These adjustments are likely unrealistic, but serve as a useful bound on the plausible magma evolution of these worlds. The last two rows of Table~\ref{Tab_Input_TRAPPIST-1} show the heating powers for both heating scenarios (reference and extreme).

Since TRAPPIST\=/1~e is closest to the star, the effect of the increased tidal heating is largest for this planet; the heating power in the high, undamped eccentricity case is larger by nearly three orders of magnitude compared to the nominal case with low, dampening eccentricity. The proximity to the host star also means that TRAPPIST\=/1 e will be most affected by additional heat sources via star-planet interactions such as induction heating \citep{Kislyakova2017}, which are not modeled here.

Figure~\ref{TR1_fluxes_volatiles} shows the influence of the extreme heating on the magma ocean evolution and the volatile budgets of TRAPPIST\=/1~e.
After an initial cooling phase, the internal heat flux is able to balance the cooling flux through the thick steam atmosphere. The mantle does not continue to cool but a magma ocean remains on the surface at the end of the simulation (after $\sim \SI{250}{\mega\year}$). In other words, a prolonged magma ocean stage is an extreme version of scenario~3, but where the surface remains hotter than 2000~K.
The prolonged magma ocean stage also affects the evolution of the volatile budgets: 
most notably, no oxygen builds up in the atmosphere.
Even though atmospheric escape produces large amounts of oxygen, the liquid magma ocean can store all of it by oxidizing \ce{FeO} to \ce{Fe2O3}. 
Furthermore, the high surface temperature prevents the formation of surface oceans. Thus, the planet does not enter evolution phase VI even after  250~Myr. Phase VI (see Fig.~\ref{Plot_TR1_scenarios_evolution}, bottom left panel) is determined by low incoming stellar irradiation that allows for condensation of \ce{H2O}, if the surface is cool enough, thus removing \ce{H2O} from the upper atmosphere and stopping atmospheric escape (moist greenhouse limit).

Due to the hot surface that prevents condensation, about 10\% of the total water content that is not dissolved in the magma ocean remains as water vapour in the atmosphere, where it undergoes continuous atmospheric escape. This atmospheric escape re-enables outgassing from the magma ocean (evolution phase III). 
In other words, the prolongation of a thick magma ocean keeps water vapour in the mantle longer, but it also leads to a prolongation of the phase with atmospheric escape driven by outgassing and, thus, counter-intuitively, to further \ce{H2O} loss from the system. 
At the same time, the prolonged presence of a thick magma ocean efficiently prevents the build-up of \ce{O2} in the atmosphere.

\subsection{Overview of TRAPPIST-1 e, f, g results}
\begin{figure*}[p]
    \centering
    \includegraphics[width=\textwidth]{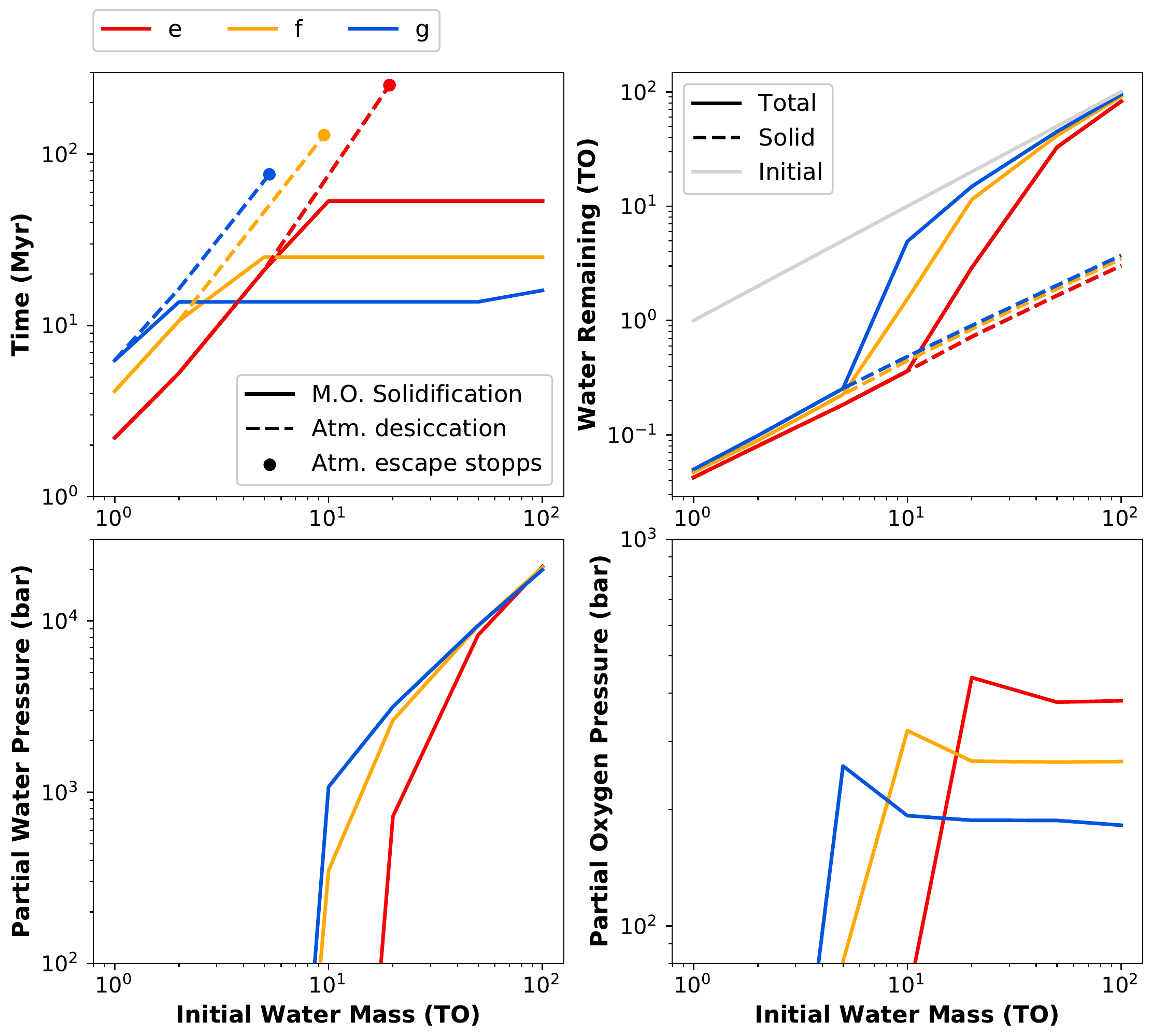}
    \caption{Final results for TRAPPIST\=/1~e, f , and g after the atmospheric escape stopped for initial water masses from $1 - \SI{100}{TO}$: \textit{Upper left:} Solidification time of the magma ocean and desiccation time of the atmosphere (i.e. atmospheric escape of water more efficient than outgassing). The dots indicate time where the escape stops (Table~\ref{Tab_Input_TRAPPIST-1}). For larger initial water masses, water will remain in the atmosphere at the end of the simulation. \textit{Upper right:} Total water remaining in the system (solid) and water locked in the solid mantle (dashed). The grey line indicates the initial water mass as comparison and upper limit. \textit{Lower left:} Partial water pressure in the atmosphere at the end of the simulation. \textit{Lower right}: Partial oxygen pressure at the end of the simulation.
    \textcolor{purple}{$\nearrow$\href{https://github.com/pbfeu/Trappist1_MagmOc/tree/master/Fig_Trappist1_Summary}{GitHub}}}
    \label{TR1_Summary}
\end{figure*}

A condensed overview of the results obtained from all our simulations are presented in Fig.~\ref{TR1_Summary}:
The upper left panel shows the mantle solidification and atmospheric desiccation time of the planets.
We define the desiccation time to be when no water is left in the atmosphere. 
However, oxygen might still be present in the atmosphere.
The time when the atmospheric escape of \ce{H2O} stops is indicated with a dot.
For small initial water masses (\eg the dry scenario~1), the planet solidifies quickly once the atmosphere is desiccated. 
In this scenario, the more water is in the system, the longer it takes to dessicate the atmosphere and thus to solidify the magma ocean.
If the initial water mass is large enough (the intermediate wet scenario~2 for more than $5-\SI{10}{TO}$), there is still a thick \ce{H2O} atmosphere present when the stellar flux has decreased such that the absorbed radiation at the location of the TRAPPIST\=/1 planets drops below the runaway greenhouse limit. 
In this case, the outgoing thermal radiation of the planets is limited to $\SI{280}{\watt\per\square\metre}$ until the atmosphere desiccates \citep[Section \ref{Sec_Flux},][]{Goldblatt2013}.
From that point on, the planets are able to cool even with a thick atmosphere.
The \ce{O2} that is building up in the atmosphere has no greenhouse effect.
This process prevents the solidification time from changing for a large range of initial water masses in scenario~2. 
It only depends on the distance of the planet to the star.
For larger initial water masses ($\ge \SI{100}{TO}$ for TRAPPIST\=/1~g), the thick steam atmosphere prolongs the lifetime of the magma oceans until they reach the limit of the runaway greenhouse, which can occur after 80~Myr for TRAPPIST-1 g and after 240~Myr for TRAPPIST-1 e, if the surface is cool enough for \ce{H2O} condensation to occur (the extremely wet scenario~3).

The upper right panel in  Fig.~\ref{TR1_Summary} shows the remaining water in the system at the end of our simulations, either when the atmosphere is desiccated or the water condenses out to form an ocean.
In addition to the total water amount, the amount of water locked into the mantle is shown. 
Furthermore, we show, as an upper limit, the initial water mass.
Up to initial water masses of $\sim \SI{10}{TO}$ (dry scenario~1), all the water in the atmosphere is lost. Only a few percent of the inital water mass remains locked in the solid mantle, where it could be outgassed via volcanism in its further evolution.
When starting with more water (intermediate scenario~2 and very wet scenario~3 for TRAPPIST\=/1~g), some will remain in the atmosphere at the end of the magma ocean stage.
We caution that, due to the logarithmic scale, it may appear that almost all water is retained for $\SI{100}{TO}$ initial water mass after the evolution, but actually 20\% of the initial water mass ($\SI{20}{TO}$) is also lost in this case. 
In absolute terms, the planets in the very wet scenarios (upper range of scenario~2 for TRAPPIST\=/1~e, f and scenario~3 for TRAPPIST\=/1~g) lose approximately the same amount of water or more than in the dryer cases. Still a much larger fraction of the initial water content remains in the planets (atmosphere and mantle) compared to the very dry scenarios 1.

The lower left panel in  Fig.~\ref{TR1_Summary} is very closely related to the upper right one: it shows the partial water pressure in the atmosphere which is left in the atmosphere when atmospheric escape of \ce{H2O} stops because water vapour does not reach the upper atmosphere anymore.

The lower right panel shows the oxygen pressure that builds up in the atmosphere due to photolysis of \ce{H2O} and the subsequent escape of hydrogen.
For initial water masses smaller than $5-\SI{10}{TO}$ (dry scenario~1), the produced oxygen can be stored completely in the mantle, bound in \ce{Fe2O3}.
For larger initial water masses (scenarios 2 and 3) up to $\SI{1000}{\bar}$ of oxygen can build up in the atmosphere.
We note that with increasing distance to the star the atmospheric escape of hydrogen is less efficient and less \ce{O2} builds up in the atmosphere.
Furthermore, for TRAPPIST\=/1~g we notice that the final oxygen pressure drops in the very wet scenario~3.
The prolonged lifetime of the magma ocean allows more oxygen to dissolve and oxidize the melt than in the intermediate scenario~2.  

\section{Discussion}
\label{sec_discuss}

\begin{figure*}[ht]
    \centering
    \includegraphics[width=\textwidth]{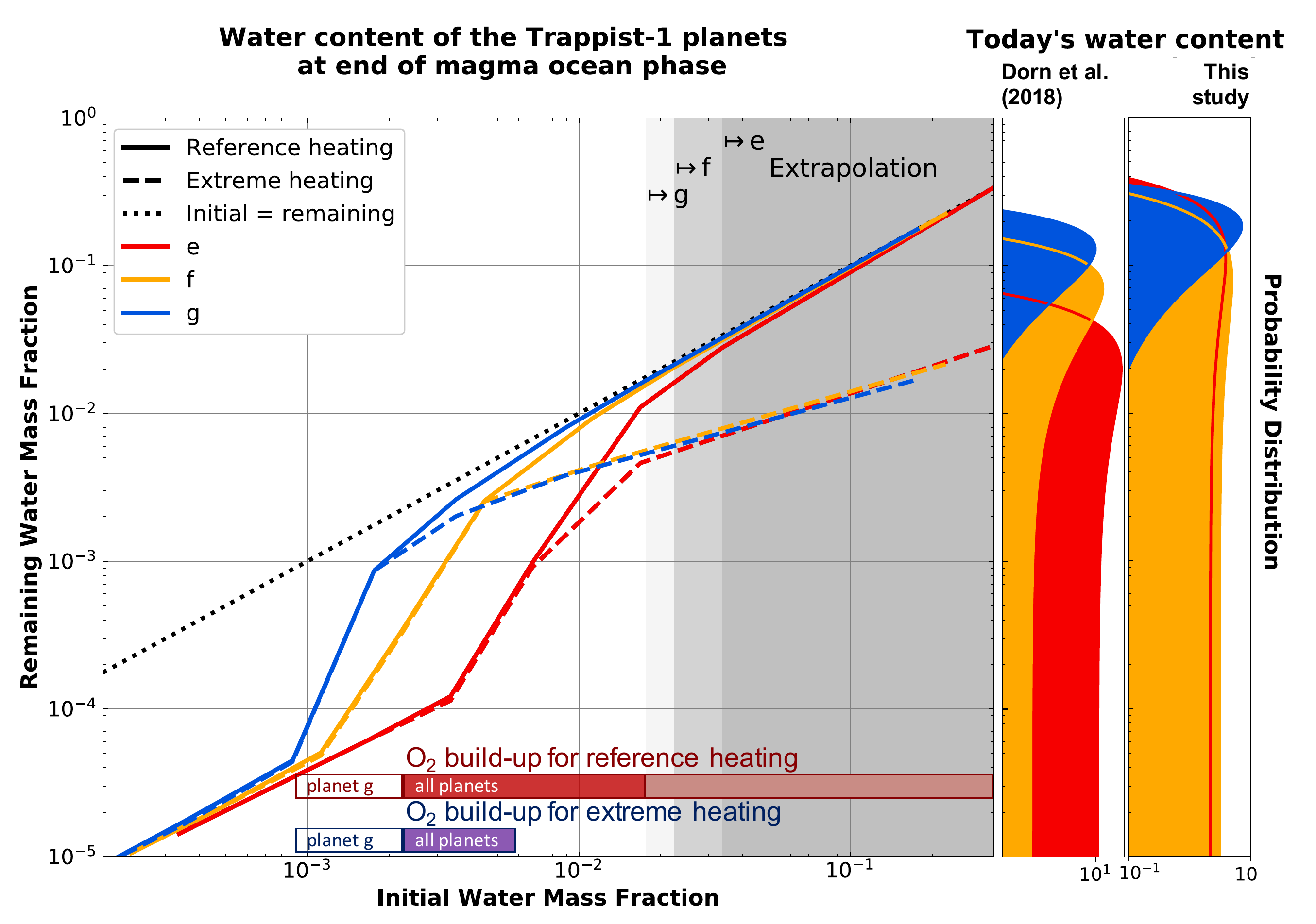}
    \caption{Final water content of the planets TRAPPIST\=/1~e, f, g with a pure steam atmosphere for different initial water mass fractions and two different heating scenarios: Reference (solid): Earth abundances of radioactive isotopes and low eccentricities; and extreme (dashed): 1000 times Earth abundance of \ce{^{40}K} and fixed eccentricities ($e=0.1$). Simulations for all planets with 100\% steam atmospheres (red, yellow, blue). Red and purple bars indicate range of initial water fractions that lead to abiotic \ce{O2} build-up. Values for initial water fractions $>20-30$~wt\% (grey area) are extrapolations on the assumption that solidification times do not increase with more water. Probability distributions show the current estimates for water content of the TRAPPIST\=/1 planets by \citet{Dorn2018} and those calculated with the interior structure model by \citet[this study]{Noack2016} with 1$\sigma$ error range for mass and radius (Table~\ref{Tab_Input_TRAPPIST-1}).
    \textcolor{purple}{$\nearrow$\href{https://github.com/pbfeu/Trappist1_MagmOc/tree/master/Fig_Trappist1_FinalWater}{GitHub}}}
    \label{TR1_Results_Oxy}%
\end{figure*}

In this work, we presented simulations of the magma ocean and volatile evolution for the planets TRAPPIST\=/1~e, f, and g. 
Studies by \citet{Dorn2018} suggest a very large water content of the TRAPPIST\=/1 planets with water mass fractions of up to 15\%, calculations with the interior structure model by \citet{Noack2016} even higher mass fractions of up to 18\%, while \citet{Agol2020} suggest a much drier composition with a water mass fraction of less than 5\%.
In Fig.~\ref{TR1_Results_Oxy}, we show an overview of the final water content of the planets TRAPPIST\=/1~e, f, and g from all our simulations:
for initial water masses from $1-\SI{100}{TO}$ and for two different heating scenarios - reference: Earth's abundance of radioactive iostopes and small, damped eccentricities; extreme: 1000 times Earth abundance of \ce{^{40}K} and fixed eccentricities ($e=0.1$).
The atmospheric composition is 100\% \ce{H2O}.

We compare our results to estimates of the current water mass by \citet{Dorn2018} and with results from the model by \citet{Noack2016} (Fig. \ref{TR1_Results_Oxy}, right).
This comparison suggests a very wet formation, i.e. with an initial water mass of more than 100 TO, especially for the planets f and g. 
As described in Sec. \ref{chap_validation}, we did not run simulations for initial water contents of more than $\SI{100}{TO}$.
For such large water masses, the water fraction in the melt becomes very large and eventually 1, which is unphysical. 
To simulate these scenarios correctly, an adaption of the physical equations would be necessary. 
To get an impression of how the magma ocean evolution would continue for larger water contents, we extrapolated the remaining water budgets up to an initial water mass of $\SI{1000}{TO}$.
We based these extrapolations on the assumption that an increase in the water content of the mantle and the atmosphere would not increase the lifetime of the magma ocean.
However, we can see already for TRAPPIST\=/1~g that $\SI{100}{TO}$ of water slightly prolongs the magma ocean lifetime and damps the build-up of oxygen.
Therefore, we can conclude that the extrapolations presented in Fig. \ref{TR1_Results_Oxy} show only upper limits.

We rather expect the extremely wet scenarios with $\SI{>100}{TO}$ to behave similar to the cases including extreme heating. For planet g, the solidification time scale rises with more than $\SI{70}{TO}$ initial water content, whereas the magma ocean solidification time remains relatively constant at about 14 Myr for $2-\SI{70}{TO}$ (see Fig.~\ref{TR1_Summary}, upper left panel). 
The extreme heating cases were found to significantly extend the magma ocean life time (to at least 250 Myr for planet e) and to prevent abiotic oxygen build-up (Fig.~\ref{TR1_fluxes_volatiles}). Further, the planet remains relatively long in a stage, where atmosphere erosion drives interior outgassing, since the surface is too hot to allow for condensation of water. This leads to an increase in the total water loss for a given planet compared to a case where the surface can solidify. With a solid surface, a water ocean can form on the surface, thus saving \ce{H2O} from being photolysed in the upper atmosphere via condensation. Testing such a hypothesis with an improvement of  our model, where we can extend the simulation beyond 250~Myr, and with a more physically plausible heating mechanism could provide further insight into habitability of the TRAPPIST-1 planets.

Even though we most likely overestimate the remaining water content in the system (especially for very wet cases), we can show that only $3-5\%$ of the initial water will be locked in the mantle due to the small partition coefficient between melt and solid. Therefore, it is not possible in our simulations to sequester large amounts of water in the mantle without several thousand bars of water in the atmosphere or at the surface. Consequently, with a water content today of several hundred terrestrial oceans and only a few percent of it being locked in the mantle, planets f and g will most likely be worlds with very deep oceans.
It should be noted that it has been postulated that instead of the magma ocean freezing from the bottom of the mantle to the top, due to the high temperature contrast between lower mantle and core, a basal magma ocean could remain during the magma ocean crystallization process \citep{Labrosse2007}.
In addition, \citet{Nomura2011} suggest that such a basal magma ocean could have extended to up to $\SI{1000}{\kilo\metre}$ thickness due to enrichment in iron, making the magma denser than the solidifying rock above. 
Such a basal magma ocean could then in theory also be a reservoir for volatiles, which, on the other hand, leads to a decrease of the melt density and might therefore destabilize such a basal magma layer.

A $\SI{300}{TO}$ water body on the surface of TRAPPIST\=/1~g would result in a $\sim \SI{670}{\kilo\metre}$ deep ocean, most likely forming high pressure ice layers in the ocean \citep{Dorn2018}. 
It is still an objective of current research to determine whether such ocean worlds are potentially habitable \citep[\eg][]{Noack2016}. 
Most notably for the further evolution of the planetary atmosphere, deep oceans may suppress further outgassing due to the high temperatures and pressures at the ocean floor even if volcanic activity is present \citep{Noack2016}, but transport through the ice layer might still be possible \citep{KalousovaSotin18}. 
Any further evolution can therefore only be driven by chemical interactions between the oxygen-rich atmosphere and the deep water ocean.
\citet{Glaser2020} suggest that, in the long-term evolution of a water world, the biotic and abiotic (by photolysis) rates of oxygen production might be very similar, further devaluing oxygen's role as a potential biosignature.
However, the masses of the TRAPPIST\=/1 planets are still debated and our results may have to be revised in the near future.

We also show that the TRAPPIST\=/1 planets start to build up several hundred bars of oxygen for initial water masses larger than $\SI{5}{TO}$. First estimates by \citet{lingam2020} indicate that once the magma ocean solidifies it would be very difficult to reduce abiotic oxygen build-up assuming present Earth sinks and sources. Though \citet{Harman2018} present that lightning might be an effective way of reducing the oxygen abundance in such an atmosphere, as long as water vapour is present in the atmosphere. We find, however, that in scenario 2 (5~TO for TRAPPIST\=/1~g, Fig.~\ref{Plot_TR1_scenarios_evolution}) it is in principle possible to build up significant amounts of free oxygen and completely erode water vapour via atmospheric escape. While such a dry oxygen-rich atmosphere is unlikely to occur for the very wet TRAPPIST\=/1~g, it could occur for the drier TRAPPIST\=/1~e. If these high oxygen pressures are a physical reality, it is possible to detect them with \textit{JWST} \citep{Lincowski2018}.

However, we also show that large heating rates can prevent the abiotic build-up of oxygen in the atmosphere and therefore could facilitate the habitability of these planets.
Heating rates of 1000 times Earth abundance of \ce{^{40}K} seem extremely high.
However, \ce{^{40}K} presents a powerful and continuous heat source for the long-term evolution (several $\SI{100}{\mega\year}$) of a rocky planet and can thus help us to understand how a planet can sustain a magma ocean over a long time or even permanently.
Also, there are further heat sources like additional tidal (\eg via planet-planet tides that were neglected here) or magnetic heating \citep[\eg][]{Kislyakova2017}, which we did not consider in this work.

Our results favor a very wet formation of the planets TRAPPIST\=/1~f and g, which provides necessary context to discuss how the delivery of such a high water mass fraction can be achieved during formation. E.g. \citet{Lichtenberg2019} show that radiogenic heating of \ce{^{26}Al} leads to desiccation in rocky planetesimals and postulates that only planet building blocks poor in \ce{^{26}Al} can lead to water-rich rocky planets. It will be needed to investigate, however, if water delivery via cometary material from the outer disk could explain a water-rich planet like TRAPPIST\=/1 g, even if the rocky planetesimals were water-poor. \citet{Miguel2020} provide dedicated formation scenarios for TRAPPIST\=/1, where they also propose that TRAPPIST\=/1~g likely formed very water-rich, TRAPPIST\=/1~e could have formed dry to water-rich, and TRAPPIST\=/1~f lies in between with respect to water mass fraction. All these scenarios compare well with our comparison between current water content estimates after magma ocean evolution and initial water content before magma ocean evolution. However, we note that the uncertainties and systematic errors that underpin current water estimates are large. Thus, more work is needed to tighten the possible ranges of water content scenarios.

One assumption that can affect the magma ocean evolution is the starting time of the simulation, i.e. the age of the star at $t = 0$.
As mentioned in Section~\ref{sec_therm_model}, we assume an age of the star of $\SI{5}{\mega\year}$.
As a late impact such as the moon forming impact on Earth can reset the magma ocean evolution, this is an important parameter to discuss.
However, the compact architecture of the TRAPPIST\=/1 system suggests that the disk was cleared of larger rocks very early in the system's evolution as a late impactor would have severely disturbed the stability of the system \citep{Gillon2017,Tamayo2017}.

So far our model assumes efficient cooling due to the strong convection of the magma ocean. This is only true until the melt fraction at the surface drops below 0.4 and the viscosity of the melt becomes solid-like.
\citet{Lebrun2013} proposed that for viscosities below 0.4 the magma ocean could enter into a `mush stage' where a thermal boundary layer (TBL) develops on the surface. A similar idea is presented by \citet{Schaefer2016}. Other models either do not take this effect into account \citep[\eg][]{Hamano2013} or stop the simulation of the solidification process at that point \citep{Elkins-Tanton2008,Nikolaou2019}. 
\citet{Debaille2009} propose that a thick thermal boundary layer, including possibly even a lithosphere, may be neglected towards the end of the magma ocean since iron-rich minerals that are the last components to crystallize from the magma ocean likely lead to overturning near the surface and a re-setting of the thermal boundary layer.

Because we aimed in this work for a versatile and numerically efficient magma ocean description, we neglected the `mush stage' towards the end of the magma ocean stage, following the argumentation of \citet{Debaille2009} and \cite{Hamano2013}. 
As \citet{Lebrun2013} note, the introduction of a TBL requires an additional iterative loop to balance the heat flux out of the TBL and the atmospheric heat flux. 
Nevertheless, it would still be interesting to investigate in a future update of the code how the volatile budget is modified towards the end of the magma ocean evolution by a TBL.

Even when taking into account the TBL, our model assumes that the convection in the mantle is described by a single Rayleigh number, which is a very simplifying assumption.
As the viscosity strongly increases with decreasing melt fraction, a layered treatment of the convection would be more accurate.
As a result, the thermal evolution and outgassing rates would slow down.
However, since the viscosity increases dramatically only for the lowest layers of the magma ocean where the melt fraction is below 0.4, we would not expect our results to change for most of the mantle solidification.
Only when the magma ocean becomes very shallow, most of the magma ocean will be at a melt fraction of around and below 0.4.
This period is represented by the `mush stage' in \citet{Lebrun2013}.

\section{Conclusion}
\label{sec: conclusion}
We investigated the magma ocean evolution of the planets TRAPPIST\=/1~e, f, and g and identified three general scenarios (Fig.~\ref{Summary_Trappist1_scenarios_g}):
\begin{itemize}
    \item Scenario~1, dry: The steam atmosphere prevents the mantle from solidifying because the absorbed stellar radiation is higher than the runaway greenhouse flux \citep[Sec.~\ref{Sec_Flux} and][]{Goldblatt2013}. Only when the planet has lost all its atmospheric water via atmospheric escape within a few Myr does the magma ocean solidify. No oxygen builds up in the atmosphere because the magma ocean efficiently removes it from the atmosphere.
    \item Scenario~2, moderately wet: The magma ocean solidifies while there is still water left in the atmosphere if the absorbed flux has dropped below the runaway greenhouse limit. At this point, the magma ocean is no longer able to remove the oxygen from the atmosphere. As the atmospheric escape of hydrogen continues, several hundred bars of oxygen can build up in the atmosphere. At the time of mantle solidification, the atmosphere will consist of free oxygen and potentially water vapour. There is also the possibility that the magma ocean ends with an oxygen-rich and dry atmosphere.
    \item Scenario~3, extremely wet:
    The thick steam atmosphere prolongs the lifetime of the magma ocean, which absorbs the photolytically produced oxygen. At the end of the magma ocean stage, the planet ends up with a thick steam atmosphere and a smaller relative contribution of oxygen.
\end{itemize}

In Scenario~3, the magma ocean phase can, in principle, be extended up to 250~Myr for TRAPPIST\=/1~e, if the planet receives continuous interior heating of $2.8 \times 10^4$~TW. The thick magma ocean prevents oxygen build-up and 90\% of \ce{H2O} remains dissolved in the magma ocean. The whole planet is, however, after 250~Myr in a state where water vapour is kept in the atmosphere due to the very hot surface (2000~K) and is eroded via photolysis. The water loss in the atmosphere is instantly replenished with efficient outgassing from the magma ocean that is still very thick (interior driven outgassing). That is, the planet still suffers continuous water loss even after 250~Myr due to the presence of a thick magma ocean. In contrast to that, without strong interior heating, the stellar irradiation would be so low and the surface so cool that water can condense at the surface, thus removing water from the atmosphere and keeping it from getting eroded.
We found consistently for all investigated planets in this work (including Earth and GJ~1132b) that only $3-5\%$ of the initial water will be locked in the mantle after the magma ocean solidified.

Comparison with the current estimates of the water content of the TRAPPIST\=/1 planets by recent studies \citep{Noack2016,barr2018interior,Dorn2018,Unterborn2018b} suggest that the potentially habitable TRAPPIST\=/1 planets have water mass fractions of 0--0.23, 0.01--0.21, and 0.11--0.24 for planets e, f, and g, respectively.
This suggests that planet~g followed scenario~3 and planet~f scenario~2 or 3, meaning that both planets are most likely covered by a thick ocean and a potentially oxygen rich atmosphere. However, significant evolution after solidification, as well as processes not included in our model, can also significantly affect atmospheric abundances. Planet~e, however, might have followed scenario~1 or 2 and could therefore be the most similar to Earth. 
The water-richness required to explain the high water content of TRAPPIST\=/1~f even after the magma ocean evolution, requires a very wet formation scenario. 
That may mean that the planetesimals from which this planet formed were poor in $\ce{^{26}Al}$ \citep{Lichtenberg2019}. The possible initial water content that is necessary to start a magma ocean evolution and still end up with the current observed water mass fraction in the TRAPPIST\=/1~e, f, and g planets are also well in line with the formation model of \citet{Miguel2020}. Those authors predict a high likelihood of water-richness for TRAPPIST-1~f after formation and that TRAPPIST-1~e could be formed dry - in line with our results for this planet when we try to reproduce the current water mass estimates by \citet{Dorn2018}.

Since only $3-5\%$ of the initial water will be locked in the mantle after the magma ocean solidified, we also conclude that for Earth the intermediate wet formation scenario regime (10 - \SI{100}{TO}) appears to be favored to derive the current low water content of 1-\SI{10}{TO}. However, we did not consider influence of the Moon impactor and other smaller impactors as \citet{zahnle2019}. 

Our work also has major implications for the habitability of the TRAPPIST-1 planets, in particular for TRAPPIST\=/1~g. $\SI{300}{TO}$ water body on the surface of TRAPPIST\=/1~g, as appears likely from the combination of formation \citep{Miguel2020}, magma-ocean evolution (this work) and current water estimates \citep{Dorn2018,Unterborn2018} would result in a $\sim \SI{670}{\kilo\metre}$ deep ocean. Such deep ocean worlds are potentially uninhabitable \citep[\eg][]{Noack2016}. Furthermore, it could build up several hundred bars of oxygen at the end of its magma ocean stage, which could be very difficult to remove \citep{lingam2020}. Thick oxygen atmospheres, however, could be detectable with \textit{JWST} \citep{Lincowski2018}, constraining evolution models further.

To conclude, \magmoc{} is a powerful tool inside the \vplanet{} code to simulate the thermal and volatile evolution of terrestrial (exo)planets during their magma ocean phase.
It is especially applicable to simulate a large parameter space for various heating rates, orbital configurations, initial compositions, and stellar types.

\section{Outlook}
\label{sec: outlook}

Even though our model successfully reproduces previous results for magma ocean evolution of Earth and the Super-Earth GJ1132b \citep{Elkins-Tanton2008,Hamano2013,Schaefer2016} and produces valuable results for the TRAPPIST\=/1 planets, there are numerous directions for future research.

The next step will be to add the powerful greenhouse gas \ce{CO2} into the model. As the example of Venus shows, \ce{CO2} can be an important constituent of atmospheres of rocky planets. 
\ce{CO2} was also a major constituent of early Earth's atmosphere \citep[\eg][]{Kasting1993a,Deng2020}.
The amount of \ce{CO2} present in the atmosphere determines whether condensation of water to an ocean can occur. 
The higher the \ce{CO2} partial pressure the more difficult it is for the water to condense \citep{Lebrun2013,Massol2016,Salvador2017,Stueken2020}.
To fully understand the early evolution of rocky planets, it is necessary to include \ce{CO2} into this model.
Our model so far can only simulate water composition up to 100~TO. Since TRAPPIST-1g has likely a higher water content than 100~TO, future work must reformulate the equation of state in a future version of \magmoc{} to coherently tackle extremely wet scenarios.

Furthermore, in this work, we assumed bulk silicate Earth composition, i.e. an oxidizing mantle composition. However, even for early Earth this assumption has come under scrutiny \citep{Kasting1993b,Gaillard2014,Ehlmann2016,Fegley2016,Schaefer2016,DelGenio2018,Armstrong2019}. Further work may be warranted to understand if and under which conditions a reducing mantle composition could occur, which would lead to outgassing of \ce{CO} and \ce{H2} instead of \ce{CO2} and \ce{H2O} as assumed here \citep{Katyal2020,Ortenzi2020}.
In addition to its effect on the redox state of the outgassed atmosphere, the mantle, and in particular the crust, composition have an important influence on the stability of liquid water on the surface after the magma ocean solidified \citep{Herbort2020}.

We invoked an extreme interior heating scenario for TRAPPIST\=/1~e via unphysically high radiogenic heating rates to explore if a magma ocean can last more than 100~Myr.  There are, potentially, more physical additional heat sources for the TRAPPIST\=/1 system, which we did not consider in the model, and which could achieve the necessary heating rate of $3\times 10^4$~TW. Due to the close proximity of planets orbiting M dwarfs to their host star, \eg electromagnetic induction heating due to the inclination of the magnetic dipole of the star \citep{Kislyakova2017} and enhanced tidal heating due to the decay of the obliquity \citep{Heller2011} could be relevant.  Also, \citet{Millholland2019} note that obliquity tides in a tight planetary system like TRAPPIST-1 could create up to $3 \times 10^3$~TW heating.

Future work will show if these additional mechanisms can indeed provide sufficient, sustained heating to lead to a long magma ocean phase, preventing the abiotic build-up of oxygen as long as the magma ocean is thick enough. In this case, we will also need to extend our model to be able to perform simulations to tackle coupled magma ocean-atmosphere evolution over Gyrs time scale. In this context, we will also need to investigate how much water will be removed from the planet, because the prolonged magma ocean presence also prolongs water erosion.

The comparison between our simulations and the current water content of the TRAPPIST-1 planets are based on composition estimates from \citet{Dorn2018}, which are in turn based on planetary radii and masses constraints of \citet{Delrez2018} and \citet{Grimm2018}, respectively, as well as updated measurements by \citet{Agol2020}. 
If new mass constraints occur that lead to new water content estimates, then these new results can be easily compared to our simulation results because we agnostically covered a large range of water abundance. 

In this work, we only simulated the planet's evolution until the magma ocean stage stops. Another obvious further line of research is to extend our model to consider outgassing from the solidified mantle or the thermal evolution of the planet's interior after the end of the magma ocean phase \citep[Garcia et al., in prep.]{Driscoll2013}. 
Then we might be able to confirm that outgassing from the mantle and crust continues even when the atmosphere was completely eroded (our scenario~1) to build up a new, potentially habitable atmosphere \citep{Godolt2019}. 
This approach would also allow predictions for the atmospheric content today that could be tested by \textit{JWST} \citep{Wunderlich2019}.
However, this would require adding the treatment of the solidified part of the mantle to the thermal evolution model.
Furthermore, taking into account the specific location of radiogenic and tidal heat production rather than using a mantle averaged approach would advance our model, as well as including the heat flux from the core into the mantle.
This is also true for the partition coefficient of water between solid and melt which we assume to be constant throughout the mantle.
However, this parameter depends on the material of the mantle and should therefore vary with depth.

We also did not take into account yet the influence of external objects like the Moon forming impactor and possible other smaller impactors that could further modify atmosphere and mantle composition as recently proposed by \citet{zahnle2019} to explain the early Earth evolution. This may be another possible avenue to expand the model.

For exoplanets, the Super-Earth 55~Cancri~e may be currently the most observationally accessible planet that has even today a magma ocean on its surface \citep{Demory2016}.The atmospheric composition of this planet is currently still debated based on tentative atmospheric \ce{HCN} observations with the \textit{Hubble Space Telescope}. \ce{HCN} would imply a reducing, \ce{N2} dominated atmosphere instead of an oxidizing atmosphere as assumed in this study \citep{Tsiaras2016,Hammond2017,Zilinskas2020}.
Future observations with \textit{JWST} or the \textit{ELT} will target 55 Cancri e as an example to study the evolution of magma ocean planets and the influence of these magma oceans on the atmosphere. Future research should reproduce the atmosphere composition of this exoplanet with an adapted model, which also takes a reducing mantle composition into account. 

Similarly, new results on Io may provide additional insight and constraints on models of magma oceans. While our work suggests TRAPPIST\=/1 e--g are not likely to possess magma oceans today, Io may still possess one \citep{Khurana11,Tyler15}. As new research on Io becomes available, it may also inform future interior modeling of the TRAPPIST\=/1 planets. In particular, NASA's \emph{Io Volcanic Observatory} \citep{McEwen14}, should it launch, will provide invaluable data in our understanding of extremely heated bodies. Future work could also apply \magmoc{} to Io for calibration and to predict features that may be detected by flybys.

The still debated composition of the TRAPPIST\=/1 planets as well as the lava world 55 Cancri e demonstrate that there is a real need to reformulate and expand geophysical models that were mainly developed for the Solar System planets, especially Earth. We have demonstrated that \magmoc{} not only has merit to simulate the magma ocean stage of Earth, shedding light on how our world became habitable, it also is versatile enough to be applicable to extrasolar planets like the TRAPPIST-1\=/ planets. For rocky exoplanets, \magmoc{} provides the ideal link between formation models such as that of \citet{Miguel2020}, interior structure studies like \citet{Noack2016,Dorn2018,Unterborn2018}, to habitability studies like that of \citet{lingam2020}, as well as to studies which explore extreme geophysical scenarios such as \citet{Kislyakova2017} and \citet{Millholland2019}. \magmoc{} is thus a valuable tool to explore the diversity of rocky words in the Solar System and beyond.

\section*{Acknowledgements}

We thank Gianluigi Ortenzi and Peter Driscoll for valuable discussion and suggestions for improvement, Andrew Linkcowski for time consuming climate simulations, and Rodrigo Luger and Hayden Smotherman for their support with \vplanet{}. 
P.B. acknowledges a St Leonard's Interdisciplinary Doctoral Scholarship from the University of St Andrews.
L.C. acknowledges support from the DFG Priority Programme SP1833 Grant CA 1795/3.
R.B.'s contribution was supported by NASA grant number 80NSSC20K0229 and the NASA Virtual Planetary Laboratory Team through grant number 80NSSC18K0829.
Th.H. acknowledges support from the European Research Council under the Horizon 2020 Framework Program via the ERC Advanced Grant Origins 83 24 28.

\section*{Author Disclosure Statement}
No competing financial interests exist.     

\bibliographystyle{aa}
\bibliography{library}

\section*{Appendix}

\begin{table}[ht]
    \caption[Values model]{Values for parameters used in \magmoc{}. If not otherwise indicated, they are taken from \citet{Schaefer2016}}
    \begin{tabular}{ccc}
    	\noalign{\smallskip}
    	\hline
    	\noalign{\smallskip}
    	Symbol & Parameter & Value \\ 
    	\noalign{\smallskip}
    	\hline \hline
    	\noalign{\smallskip}
    	$\rho_\mathrm{m}$  ${}^{a}$ & Mantle bulk density& $\SI{4000}{\kilogram\per\cubic\metre}$\\
    	$c_p $ & Silicate heat capacity & $1.2 \times 10^3 \si{\joule\per\kilogram\per\kelvin}$ \\
    	$\Delta H_\mathrm{f}$ & Heat of silicate fusion & $4 \times 10^5 \si{\joule\per\kilogram}$ \\
    	$\alpha$ & Thermal expansion coefficient & $2 \times 10^{-5} \si{\per\kelvin}$ \\
    	$k_{\ce{H2O}}$ & Water part. coeff. melt - solid & $0.01$\\
    	\noalign{\smallskip}
    	\hline
    \end{tabular}
    \\
    ${}^{a}$ \citet{Lebrun2013}
    \label{Tab_Therm_Model_Value}
\end{table}

\end{document}